\documentclass[reqno,11pt]{amsart}
\usepackage{amsmath, amssymb, amsthm,amsfonts, bm} 
\usepackage[english]{babel}
\usepackage{bbm}
\usepackage{graphicx}
\usepackage{url}
\usepackage{epstopdf}
\usepackage[a4paper,bindingoffset=0.5cm,left=2cm,right=2cm,top=2.5cm,bottom=2cm,footskip=.8cm]{geometry}
\usepackage[pdfencoding=auto]{hyperref}
\usepackage{kbordermatrix}
\usepackage{tikz}
\usetikzlibrary{arrows, automata,positioning,calc,shapes,decorations.pathreplacing,decorations.markings,shapes.misc,petri,topaths}
  \usepackage{pgfplots}
  \pgfplotsset{compat=newest}
  \usetikzlibrary{plotmarks}
  \usepackage{grffile}
\newlength\figureheight
  \newlength\figurewidth
\pgfplotsset{%
    tick label style={font=\scriptsize},
    label style={font=\footnotesize},
    legend style={font=\footnotesize},
         every axis plot/.append style={very thick}
}
 
\usepackage{rotating}
\usepackage{amsbsy,enumerate}
\usepackage{graphicx}
\usepackage{ccaption}
\usepackage{comment}
\usepackage{mathrsfs} 
\usepackage{mathtools}
\usepackage{slashbox}
\usepackage{xcolor}
\usepackage[normalem]{ulem}
\usepackage[linesnumbered,ruled,vlined]{algorithm2e}
\usepackage{subfig}

\makeatletter
\newcommand{\specialcell}[1]{\ifmeasuring@#1\else\omit$\displaystyle#1$\ignorespaces\fi}
\makeatother

\newcommand{\vb}{\vspace{3.2mm}}

\allowdisplaybreaks

\theoremstyle{plain}
\newtheorem{theorem}{Theorem}

\newtheorem{lemma}{Lemma}
\newtheorem{proposition}{Proposition}

\theoremstyle{definition}
\newtheorem{definition}{Definition}

\title[Social reinforcement learning]{How social reinforcement learning can lead to metastable polarisation and the voter model}

\author{BV Meylahn and  JM Meylahn}

\begin{document}
\begin{abstract}

Previous explanations for the persistence of polarization of opinions have typically included modelling assumptions that predispose the possibility of polarization ({i.e., assumptions allowing a pair of agents to drift apart in their opinion such as repulsive interactions or bounded confidence}). An exception is {a} recent {simulation study} showing that polarization is {persistent} when agents form their opinions using social reinforcement learning. 
{Our goal is to highlight the usefulness of reinforcement learning in the context of modeling opinion dynamics, but that caution is required when selecting the tools used to study such a model.}
We show that the polarization observed in the model of the simulation study {cannot persist indefinitely}, and exhibits consensus asymptotically with probability one. By constructing a link between the reinforcement learning model and the voter model, we argue that the observed polarization is metastable. Finally, we show that a slight modification in the learning process of the agents changes the model from being non-ergodic to being ergodic. 
Our results show that reinforcement learning may be a powerful method for modelling polarization in opinion dynamics, but that the tools {(objects to study such as the stationary distribution, or time to absorption for example)} appropriate for analysing such models crucially depend on the{ir} properties {(such as ergodicity, or transience)}. {These properties} are determined by the details of the learning process {and may be difficult to identify based solely on simulations}.

\vb

\noindent
{\sc AMS Subject Classification (MSC2020).} Primary: 91-10 (Mathematical modeling or simulation for problems pertaining to game theory, economics, and finance), 91A22 (Evolutionary games); Secondary: 91D15 (Social learning)
\vb

\noindent
{\sc PACS number(s).} 02.50.Le (Decision theory and game theory), 87.23.Ge (Dynamics of social system), 89.75.Fb (Structures and organization in complex systems)
\vb

\noindent
{\sc Affiliations.} BV Meylahn is affiliated with: Korteweg-de Vries Institute for Mathematics, University of Amsterdam; Science Park 904, 1098 XH Amsterdam; The Netherlands ({\it contact}: {\tt\scriptsize  b.v.meylahn[at]uva.nl}).
JM Meylahn is affiliated with: Department of Applied Mathematics, University of Twente; Enschede, The Netherlands

\vb

\noindent
{\sc Acknowledgments.} This research was supported by the European Union’s Horizon 2020 research and innovation programme under the Marie Skłodowska-Curie grant agreement no. 945045, and by the NWO Gravitation project NETWORKS under grant no. 024.002.003. \includegraphics[height=1em]{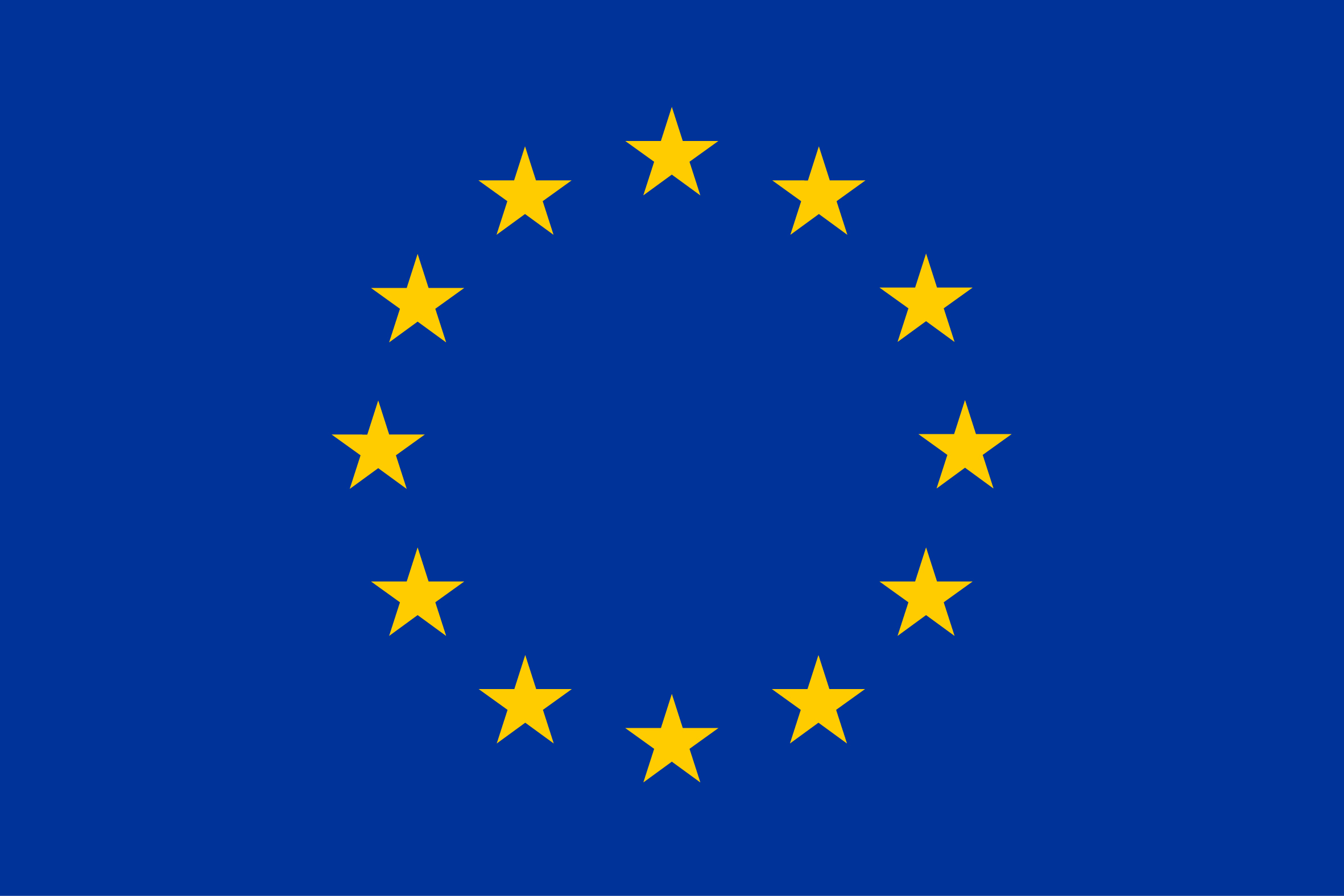}. 
The authors express their gratitude to Sven Banisch, Michel Mandjes and Hugo Touchette for their thoughtful comments and feedback on early drafts of this paper. Version: \today.
\vb

\end{abstract}

\maketitle
\section{Introduction}
Since at least 1964 scientists have been trying to answer the question ``what on earth must one assume to generate the bimodal outcome of community cleavage studies''~\cite[p.\ 153]{Abelson1964}.
Possible answers to this question have been presented; bounded confidence~\cite{Deffuant2000,Hegselmann2002, Weisbuch2004, Serrano2012} whereby agents stop listening to others if their opinion is too different from their own, repulsive forces between agents~\cite{Flache2011, Sobkowicz2012, Altafini2013, Chan2022, Burke2022} based on possible negative connections within a network or messages eliciting the opposite effect within a recipient,
stubbornness of an agent toward changing their opinion~\cite{Flache2004, Galam2011,Yildiz2013}, and distinguishing between an agent's expressed opinion and their internal opinion~\cite{Gaisbauer2020, Searle2023}. What unifies these explanations is that the resulting models all include some element from which one might infer the possibility of polarization. {Models with only attractive forces on the other hand, typically lead only to consensus (see for example the models of weighted averaging~\cite{French1956,Harary1959,degroot1974,Chatterjee1977} and imitation~\cite{Clifford1973,Holley1975}).} {Thus it comes as a surprise that the model studied by Banisch and Olbrich~\cite{Banisch2019} with only attractive forces between agents, seems to exhibit persistent polarization.}

{The dynamics in models of opinion formation typically take place on a network. A network consists of nodes (representing agents) and edges between them (representing social influence or ties). Nodes that share an edge are said to be neighbours in the network. A well studied class of opinion dynamics models on networks from the sociophysics literature is the class of voter models~\cite{Clifford1973,Holley1975} (see~\cite{Castellano2009} for an introduction). In these models a random agent is selected each round to update their opinion. The agent does this by copying the opinion held by one of their neighbours. Reinforcement learning is a model for learning by feedback: actions (or opinions in our case) for which an agent receives positive feedback are reinforced. Actions that receive negative feedback on the other hand are less likely to be taken in the future.}

{For convenience, in the remainder we refer to the paper of Banisch and Olbrich~\cite{Banisch2019} as BO while we refer to the reinforcement learning model they study as the `Asymmetric Reinforcement Learning for Opinion Dynamics model,' or simply the ARLOD model. }
{This} influential model includes no {repulsive} element {in the interaction between agents}. It proposes modelling the evolution of opinions {(of agents on connected networks)} using multiagent reinforcement learning, where agents interact via a coordination game. They find, using simulations, that allowing agents to learn their opinion through trial and error gives rise to the emergence of {persistent}\footnote{In the original article BO~\cite{Banisch2019}, `persistent' and `stable' are used interchangeably. In order to avoid confusion, we use `persistent' to discuss their claims about the ARLOD model and `stable' when making our own claims.} polarization. This is surprising, {because in this model after an interaction between two agents, the opinions of the two necessarily get closer together and cannot remain unchanged or get further apart. That is, in the {ARLOD} model there are no repulsive forces or assumptions of bounded confidence}\footnote{Models of opinion dynamics may be classified into `assimilative,' `repulsive' and models with `similarity bias'~\cite{Flache2017}. The model under consideration here does not traditionally fall in the category of models with only assimilative forces between agents because it utilizes experience based learning.}. T\"ornberg \textit{et al.}\,\cite{Tornberg2021} build on the ideas of the {ARLOD model} by incorporating the role of agent identity. T\"ornberg~\cite{Tornberg2022}, similarly looking for drivers of polarisation without the assumption of negative influences but dissatisfied with BO's assumption of selective exposure (a fixed and constant network), analysed a model which includes non-local interaction to model the effect of media. 
A variant of the reinforcement learning model with multiple opinions and synchronous updating has been studied in~\cite{Yu2016}. Their results highlight the difficulty of reaching consensus in complex networks using reinforcement learning. The idea of modelling opinion dynamics by reinforcement learning has been built on since (e.g.~\cite{Chen2020,Lorenz2021, Botte2022,Lefebvre2024}).


{An overarching goal in this paper is to highlight the importance of the relationship between model assumptions and characteristics. It can be tempting to design a model and study its characteristics by simulation. However, to accurately present the results of such a simulation study it may be important to first identify certain model characteristics. We illustrate the importance of this by presenting three results on the ARLOD model by BO~\cite{Banisch2019}.}

We show analytically that consensus is reached in the {ARLOD model} with probability one in the long run. {The} polarisation found in~\cite{Banisch2019} necessarily gives way to consensus eventually. To elucidate this result, we run simulations to estimate the tail probabilities for the time to consensus. We find that these exhibit heavy tails, indicating that there may be metastable states (corresponding to polarization) in which the model resides for a long time before reaching consensus. 

The phenomenon of metastable polarization together with eventual consensus has previously been observed in the context of the voter model, first introduced in~\cite{Holley1975}. Specifically, consensus is reached eventually if the state space of the model is finite~\cite{Cox1989}, and it has been shown that polarisation is metastable in certain network topologies~\cite{Suchecki2005a,Suchecki2005b,Castellano2003,Vilone2004}. Recently, metastable opinion polarisation has been identified in~\cite{Banisch2023} where it is shown to arise from biased information processing.

The dynamics of the voter model on networks consists of agents adopting one of their neighbours' opinions at random. At first sight, dynamics of this kind seem rudimentary in comparison to the sophisticated dynamics of reinforcement learning. However, we show that, under a separation of time scales, the {ARLOD model} converges in distribution to a voter model. This relationship highlights that the polarisation observed in the {ARLOD model} may indeed be \textit{metastable} depending on the network structure. It also bridges the seemingly disparate approaches to modelling opinions: sociophysics and computational sociology. {These two approaches differ in their typical level of abstraction, and whether they aim for tractability by keeping the dynamics simple or aim to approach realism by modelling the agents with a relatively high level of sophistication. The ARLOD model falls in the class of computational sociology seeing as the agents in the model are sophisticated enough to learn from experience. The relationship we show between this model and the voter model (a very simple model where agents imitate one another) is thus an example of a bridge between the two approaches to studying polarization.}

In designing their model, BO~\cite{Banisch2019} {decide to make} the interaction-learning relationship asymmetric{:} only one of the agents partaking in the interaction is allowed to explore and learn from the experience. We show that adapting the model to be symmetric fundamentally changes the nature of the opinion dynamics from being non-ergodic to being ergodic. Under this model, consensus is no longer absorbing so that the tools appropriate for studying polarization and consensus differ from those required in the case of the {ARLOD model}. {For example, in an ergodic system the stationary distribution may be estimated by studying the mean return time to polarized (or consensus) states. On the other hand for a non-ergodic system with absorbing states one typically studies the time to absorption (in a consensus or polarized state if these are indeed absorbing) or the number of visits to transient states before absorption. If there are both consensus and polarized absorbing states, the relative probability of consensus or polarization can be studied.}

\section{Results}\label{sec:results}

In this section, we present the asymptotic analysis of the asymmetric reinforcement learning for opinion dynamics (ARLOD) model presented by Banisch and Olbrich~\cite{Banisch2019} in the long-time limit, its relation to the voter model and the asymptotic analysis of a symmetric modification of the model.

All three analyses {(on the ARLOD model, the symmetrized version thereof, and the relationship between the ARLOD model and the voter model)} consider the same reinforcement learning method, namely, Q-learning. By using Q-learning, agents assign an estimate of the ``quality'' of expressing each opinion to a randomly selected neighbour called a Q-value. 
We present the ARLOD model for completeness of the current text. We refer to this model as the \textit{asymmetric} model because in the interaction between two agents the roles are distinguishable. One agent is chosen to express their opinion to another, who merely responds. Only the first agent updates their Q-values, and only the first agent can explore.

{Different notions of stability exist in various fields related to the model we study. To avoid confusion we present the definition of a stable state as used in this and the Methods section of the paper. The notion we use is strongly related to the notion of absorption. In the Introduction and Discussion sections we revert to using `absorbing' and `stable' separately as we discuss these notions outside of the paradigm of the model we study here.
\begin{definition}[Stability]\label{def:stable}
    A state (or class of states) is stable if once the process has entered this state (class of states), it remains there indefinitely. 
\end{definition}
}

\subsection{Asymptotic behaviour}
\subsubsection{{The asymmetric reinforcement learning opinion dynamics (ARLOD) model}}
This model of learning through social feedback considers $N \in \mathbb{N}$ agents on a random (connected) geometric network topology~\cite{Dall2002}. In particular, the network is given by $G=(V,E)$ where $V$ are the vertices representing agents, and $E$ are the connections between agents. The graph is constructed according to the random geometric graph model with radius $r_g$ (for details, see \S\ref{sec:method_graph} and Appendix~\ref{app:geograph}). Initially, all agents $i\in\{1, \ldots, N\}$ assign a (possibly random) quality $Q^{i}_{o}\in[-1,1]$ to each opinion\footnote{Note that in the simulation we initialize these values in $[-0.5,0.5]$ instead of $[-1,1]$ which is all that is required for the theoretical analysis. We do this following BO's original simulation. The reason provided is to have on average half the agents favouring each opinion.} $o\in\{-1, 1\}$. An agent holds the opinion which they assign the higher quality. In each discrete time step $t$, an agent $i$ is chosen uniformly at random to express their opinion $o_i(t)$ to a randomly selected neighbour $j$. This neighbour responds by either punishing them if the expressed opinion differs from their own ($R_j=-1$), or rewards them if the expressed opinion is shared ($R_j=1$).

Agents thus learn the value of each of the two possible opinions $\{-1,1\}$ from their experiences using stateless Q-learning. {This means that} each opinion $o$ is assigned a Q-value $Q_{o}$, measuring its ``quality'', which is updated as follows for the opinion $o_{i}(t)$ expressed in round $t$:
\begin{equation}\label{eq:Q_update}
    Q^i_{o_{i}(t)}(t+1) = Q^i_{o_{i}(t)}(t)(1-\alpha) + R_j\alpha.
\end{equation}
Here $\alpha\in (0,1)$ is called the learning rate. The Q-value of the opinion they did not express is not altered so that
\begin{equation}
    Q^i_{-o_{i}(t)}(t+1) = Q^i_{-o_{i}(t)}(t).
\end{equation}
We assume that the agent chosen to express their opinion exploits their favoured opinion (the one with the greater Q-value) with probability $1-\epsilon$ and explores by expressing their disfavoured opinion with probability $\epsilon>0.1.$ {This is known as $\epsilon$-greedy Q-learning with fixed exploration rate $\epsilon.$} 

\begin{figure}
    \centering
    \includegraphics[scale = 0.7]{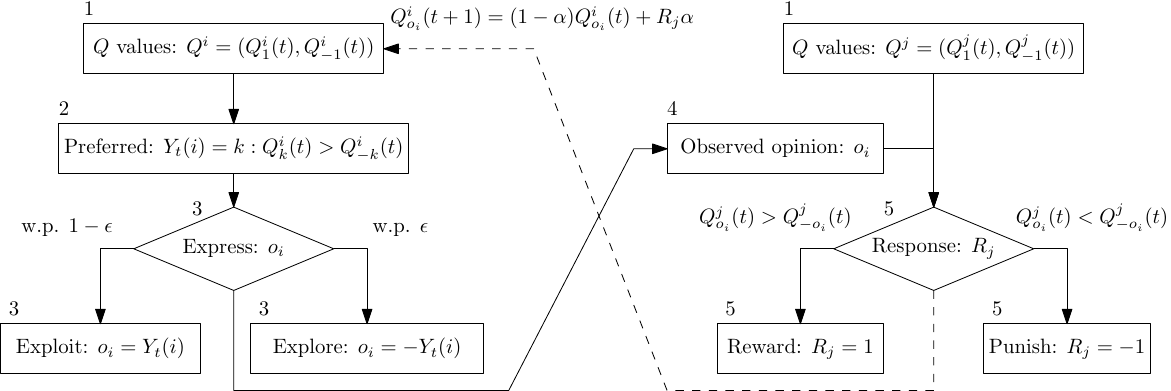}
    \caption{A schematic of the procedure followed by the two agents selected to interact in one round of the ARLOD model, as originally described in~\cite{Banisch2019}. Agent $i$ expresses an opinion $O_i$ to their neighbour $j$, who responds by punishing or rewarding agent $i$. Agent $i$ updates the Q-value for the opinion they expressed accordingly. {The numbers to the top left of the boxes indicate the suggested order for reading the schematic.}} 
    \label{fig:QL_scheme}
\end{figure}

The dynamics per round are depicted in a schematic in Figure~\ref{fig:QL_scheme}. Note that only agent $i$ adjusts their Q-values after such an interaction, and that agent $j$'s response is deterministic (honest).

\subsubsection{{Asymptotic consensus and non-ergodicity}}
We now prove that in the ARLOD model the long-time limit of the dynamics necessarily results in consensus and does not allow for polarization. The proof is inspired by the proof of an analogous result for agents who learn by simple exponential smoothing in~\cite{Meylahn2024}. We explore the time to consensus by means of simulation. For the details on the simulation, see \S\ref{sec:sim_settings}.
\subsubsection*{Analytical results}
Our first result states that consensus is a stable state. In this regard, we define consensus as the state of the model in which the Q-values each agent assigns to the opinions have the same ordering. Note that we use a slightly different notation to that used by BO. We define $Q_{o}^i(t)$ as the Q-value that agent $i\in \{1,2,\ldots,N\}$ assigns to opinion $o\in\{-1,1\}$ at time $t\in \mathbb{N}$.
\begin{lemma}[Consensus is stable]\label{lem:stay}
    If there exists a time $t_0$ such that $Q_{o}^i(t_0)>Q_{-o}^i(t_0)$ for some opinion $o\in\{-1,1\}$, and each agent $i\in \{1,2,\ldots,N\}$, then $Q_o^i(t)>Q_{-o}^i(t)$ for all $t\geq t_0$ and for all agents $i\in \{1,2,\ldots,N\}$.
\end{lemma}
We prove Lemma~\ref{lem:stay} in Appendix~\ref{sec:proofstay}. The proof follows from the fact that agents respond honestly, so that once all agents have the same ordering of Q-values, each exploration is punished while each exploitation is rewarded. This preserves the Q-value ordering.  

The next result required to prove that consensus is reached with probability one in the long-time limit, is that consensus is reachable from any state that is not consensus. 
\begin{lemma}[Consensus is reachable from all other states]\label{lem:reach}
    If the learning rate $\alpha>0$, the exploration rate $\epsilon>0$, and {$G$ is connected} then the probability of reaching consensus in finite time is positive, i.e.,
    \begin{equation}
        \mathbb{P}(\exists t_1<\infty: Q_o^i(t_1)>Q_{-o}^i(t_1), \forall i\in\{1,\ldots,N\})>0,
    \end{equation}
    for some $o\in \{-1,1\}$.
\end{lemma}
Lemma~\ref{lem:reach} is proved in Appendix~\ref{sec:proofreach} and hinges on the realisation that the ordering of an agent's Q-values may switch in a finite number of rounds as long as they have a neighbour whose Q-value ordering differs from theirs. The number of rounds required for this switch to occur is bounded from above by $2r+2$ with
\begin{equation}\label{eq:r_switch}
  r=\bigg\lceil \frac{\log (\xi)}{\log  (1-\alpha)}\bigg\rceil,
\end{equation}
for some $\xi\in(0,\alpha)$. Note that Lemma~\ref{lem:reach} is true for all connected graphs between $N<\infty$ agents and all starting states (Q-values of agents) that are not in consensus. Furthermore, consensus on either of the two opinions is reachable in this way. 

We now state the first main theorem of the paper, which states that consensus is reached with probability one in the long run in the ARLOD model.
\begin{theorem}[Consensus is guaranteed]\label{thm:con_g}
    If the learning rate $\alpha\in(0,1)$, the exploration rate $\epsilon\in(0,1)$, {and $G$ is connected} then the probability of consensus in the long run is one, i.e.,
    \begin{equation}
    \mathbb{P}(\exists t_0<\infty: Q_o^i(t)>Q_{-o}^i(t), o\in\{-1,1\} ,\forall i\in\{1,\ldots,N\}, \forall t\geq t_0) =1.
    \end{equation}
\end{theorem}
\begin{proof}
By Lemma~\ref{lem:stay} consensus is stable and so once it is reached it persists. By Lemma~\ref{lem:reach} the probability of reaching consensus from not consensus in $R=(N-1)(2r+2)$ rounds is bounded from below by $p>0$. Thus, the probability of not reaching consensus in $kR$ rounds is bounded from above by
\begin{equation}\label{eq:bound_nabsorb}
    \mathbb{P}(\nexists t_0\leq kR : Q_o^i(t)>Q_{-o}^i(t), o\in\{-1,1\} ,\forall i\in\{1,\ldots,N\}, \forall t\geq t_0) \leq (1-p)^k.
\end{equation}
The probability of never being absorbed is then bounded from above by the limit of (\ref{eq:bound_nabsorb}) as $k\to \infty$ which is zero. Therefore, the probability of the complement is one.
\end{proof}

This implies that the polarisation observed as {persistent} in the presentation of the original model's simulation {cannot persist indefinitely}. In particular, the probability reported in Figure~5 of BO should be reinterpreted from `probability of consensus' to `probability of consensus before time $N\times 20\,000$.' Furthermore, this implies that the probability of the system being in a polarised state tends to zero as $t\to\infty.$

{Note that the conditions on the network are only that it is connected. This is not a significant limitation. Studying polarization is most interesting in connected networks where there is still interaction between agents that disagree. The results also hold separately for each component of a disconnected network. Though consensus within each component does not imply consensus between components.}


\subsubsection*{Simulations}
In light of Theorem~\ref{thm:con_g}, we investigate the time to consensus as a function of the radius of the geometric network structure by simulation. The parameter settings are stated and motivated in \S\ref{sec:sim_settings}. We define the time to consensus $\tau$ as
\begin{equation}
    \tau:= \min \{t:Q_o^i(t)>Q_{-o}^i(t), o\in \{-1,1\},\forall i\in \{1, \ldots, N\}\}.
\end{equation}

\begin{figure}[htb]
    \centering
    \subfloat[]{\includegraphics[scale = 0.75]{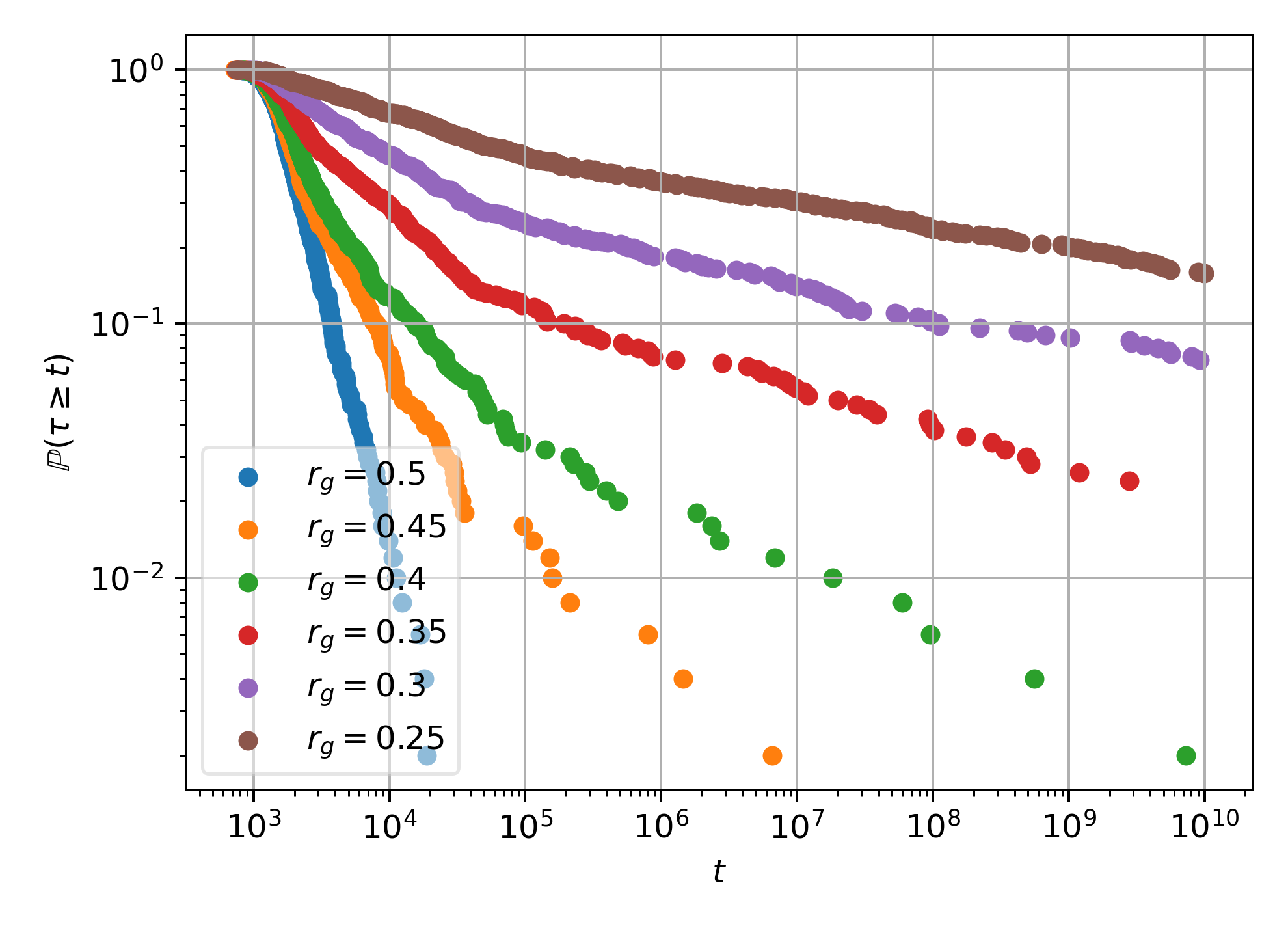}\label{fig:con_tail}} 
    \subfloat[]{\includegraphics[scale = 0.75]{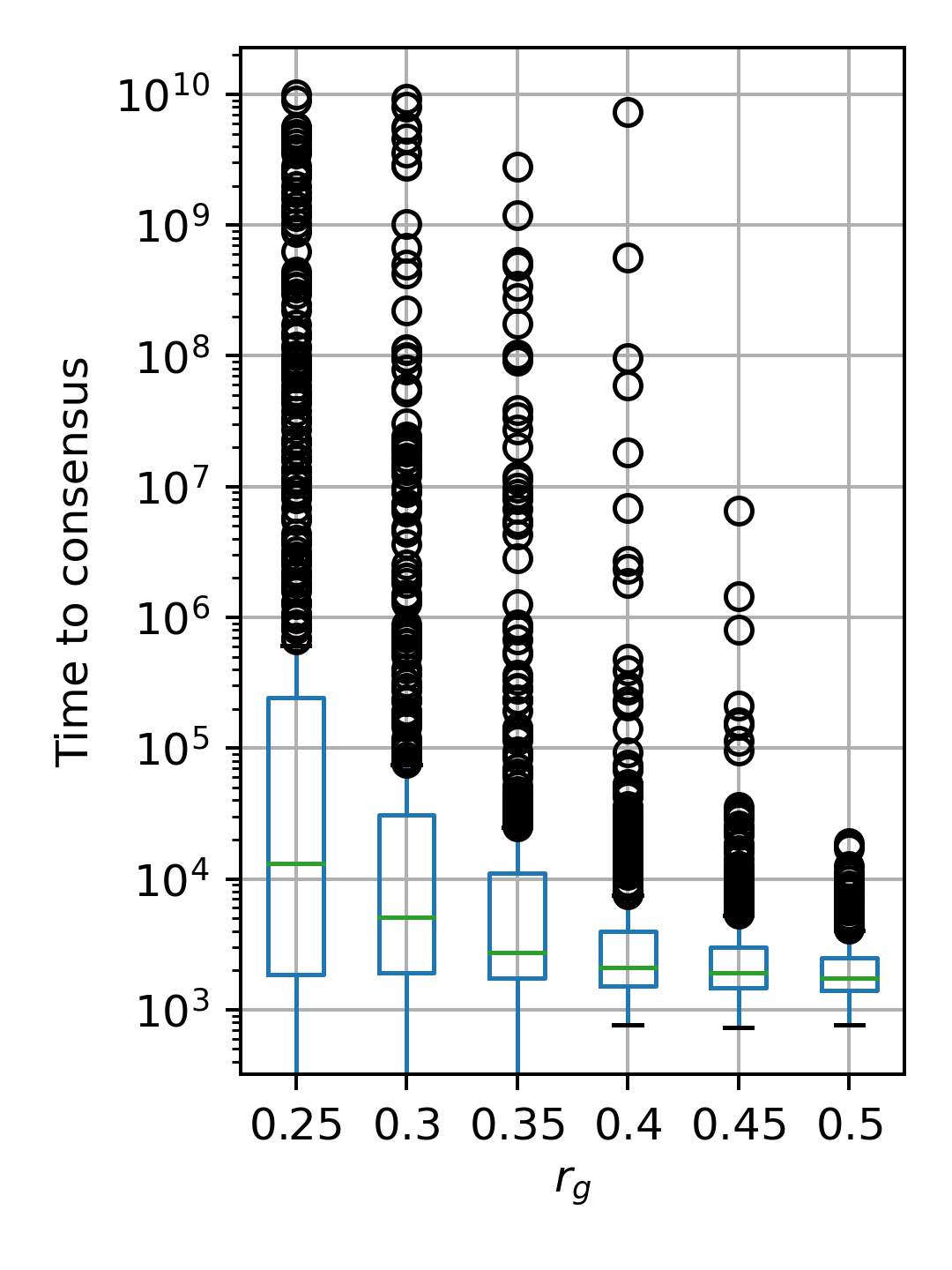} \label{fig:box}}
    \caption{{(A)} Tail probabilities ($\mathbb{P}(\tau>t)$) (on a log-log scale) and {(B)} a box and whisker diagram for the time to consensus for different values of the radius $r_g$ used in the random geometric graph model to sample networks. The linear nature of these plots are indicative of a heavy tailed distribution. The high number of outliers on the upper end of the time to consensus is indicative of a heavy-tailed distribution. The parameter settings are detailed in \S\ref{sec:sim_settings}.}
\end{figure}

In Figure~\ref{fig:con_tail}, we show the tail probabilities of the time to consensus $\mathbb{P}(\tau\geq t)$ for different radii of the random geometric graph model on a logarithmic scale. A clear pattern emerges; the bigger the radius, the sooner consensus is reached. We also note that the distributions exhibit heavy tails, especially for the smallest three settings of the radius: $r_g \in\{0.25, 0.3, 0.35\}$. This can be seen by the near linear lines (on the log-log scale) which are representative of power-law and log-normal distributions.

In Figure~\ref{fig:box}, we show box and whisker diagrams of the simulated time to consensus (conditioned on $\tau<t_{\text{max}} = 10^{10}$). This representation of the simulated data clearly shows that there are many runs which might be identified as `outliers.' This indicates that the time to consensus has a high skewness and, like the tail probabilities, points towards a heavy-tailed distribution. A possible explanation for the heavy-tails is {the existence of} metastable states, which the system may spend a lot of time in before eventually `jumping' out to consensus. Indeed, similar heavy-tailed survival probabilities were observed for the voter model on small-world networks, which exhibit metastable polarisation~\cite{Castellano2003}. We see that as the radius $r_g$ decreases, the probability that consensus is reached after time $t\in \mathbb{R}$ increases. This shows how quantitatively the dynamics do depend on the realisation of the network structure.

\begin{figure}[p]
    \centering
    \includegraphics[width =0.99\textwidth]{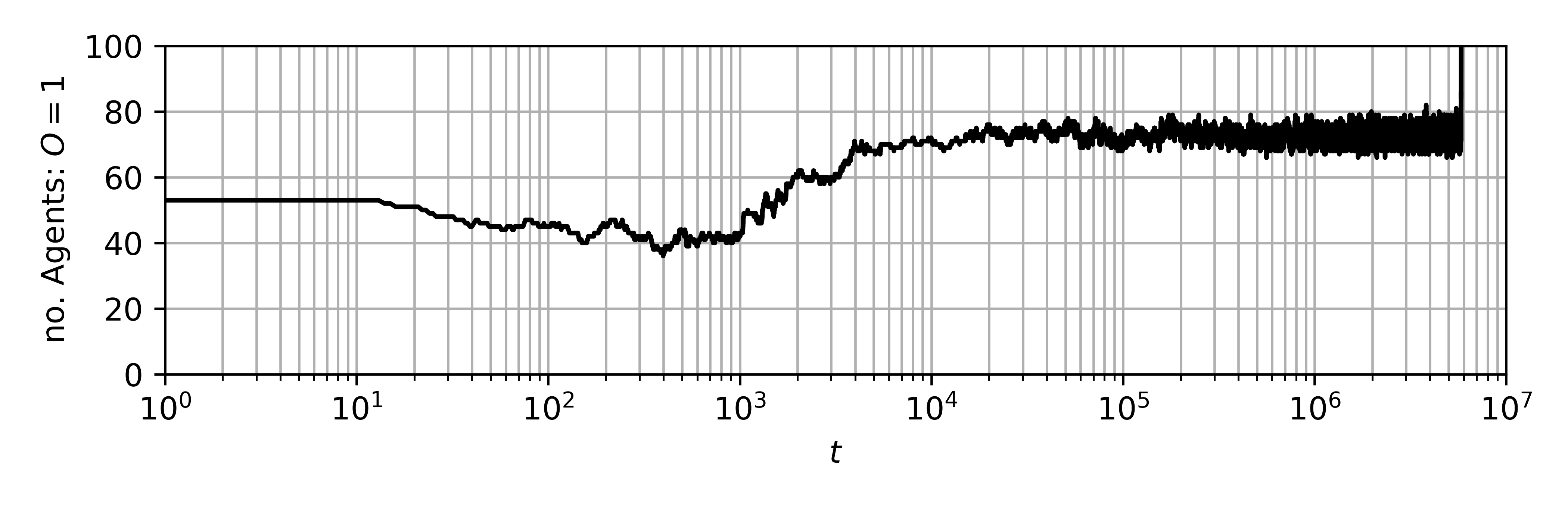}
    \caption{Number of agents holding opinion $o=1$ in a simulation run exhibiting metastable behaviour plotted with time on a logarithmic scale. The state of the network is plotted for telling timestamps of this simulation run in Figure~\ref{fig:ms_nets}. In this simulation run $r_g=0.25$, the other parameters are as in \S\ref{sec:sim_settings}.}
    \label{fig:ms_total_log}
\end{figure}

\begin{figure}[p]
    \centering
    \subfloat[]{\includegraphics[trim = {0 0 65 0}, clip = true, width = 0.27\textwidth]{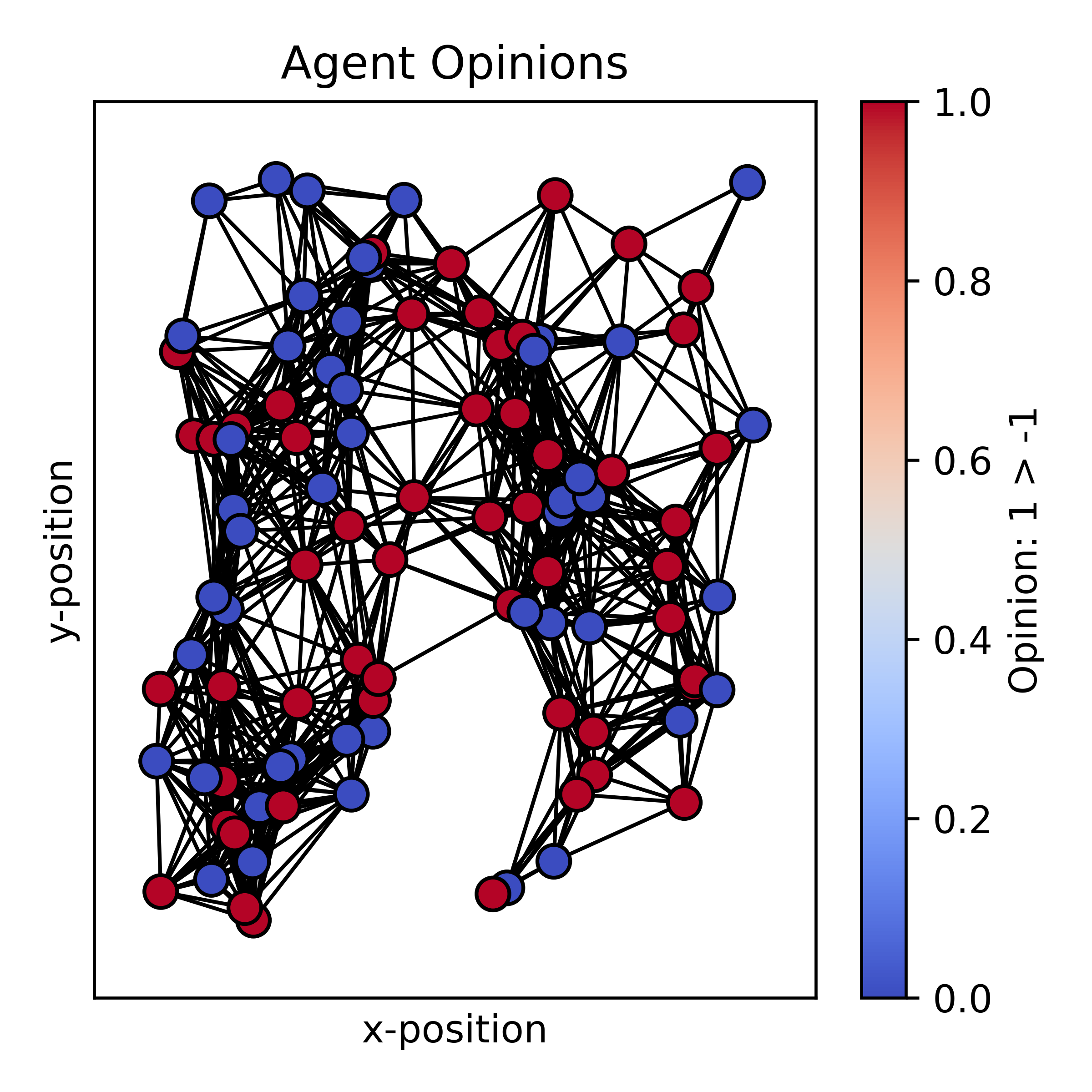}}
\subfloat[]{\includegraphics[trim = {0 0 65 0}, clip = true, width = 0.27\textwidth]{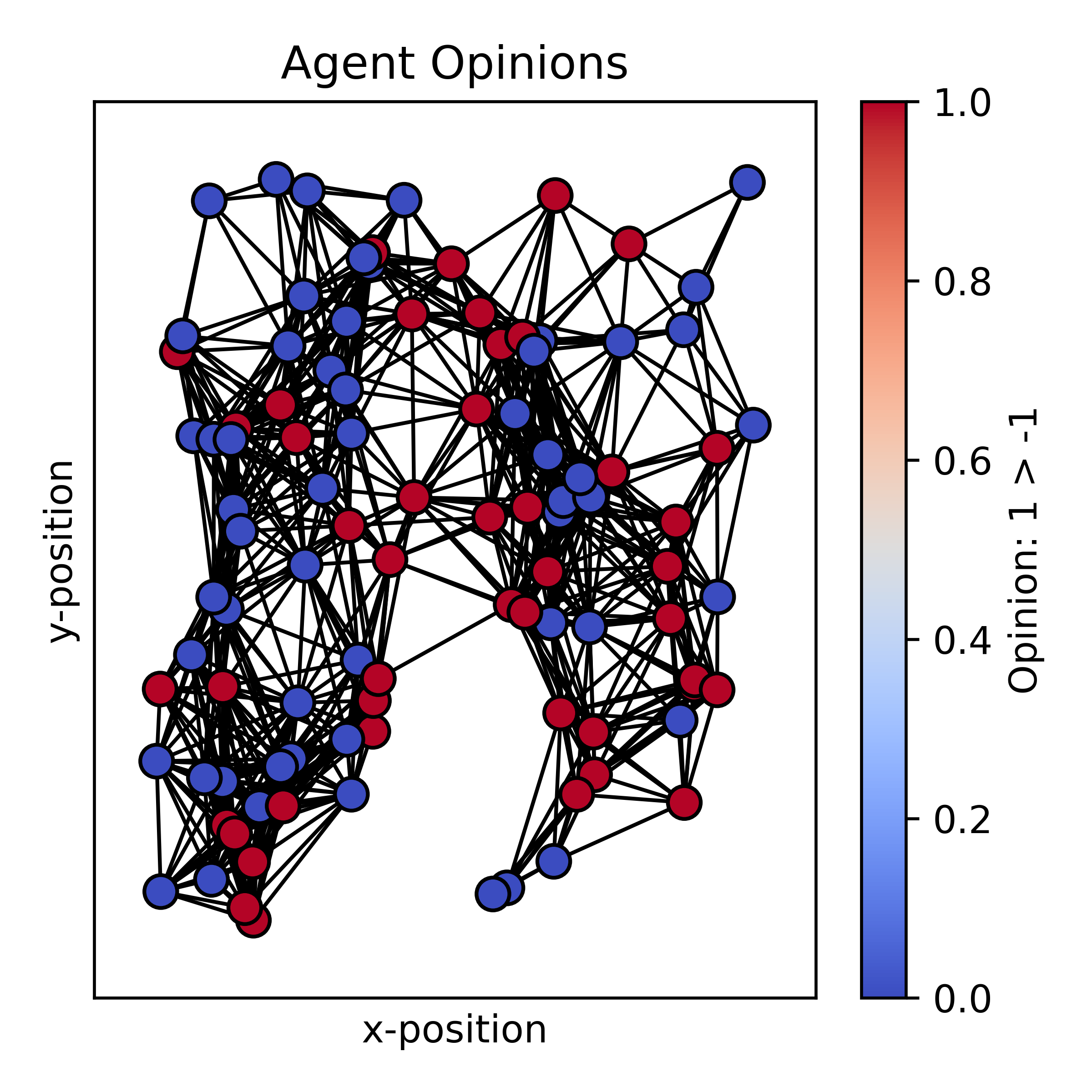}}
\subfloat[]{\includegraphics[trim = {0 0 65 0}, clip = true, width = 0.27\textwidth]{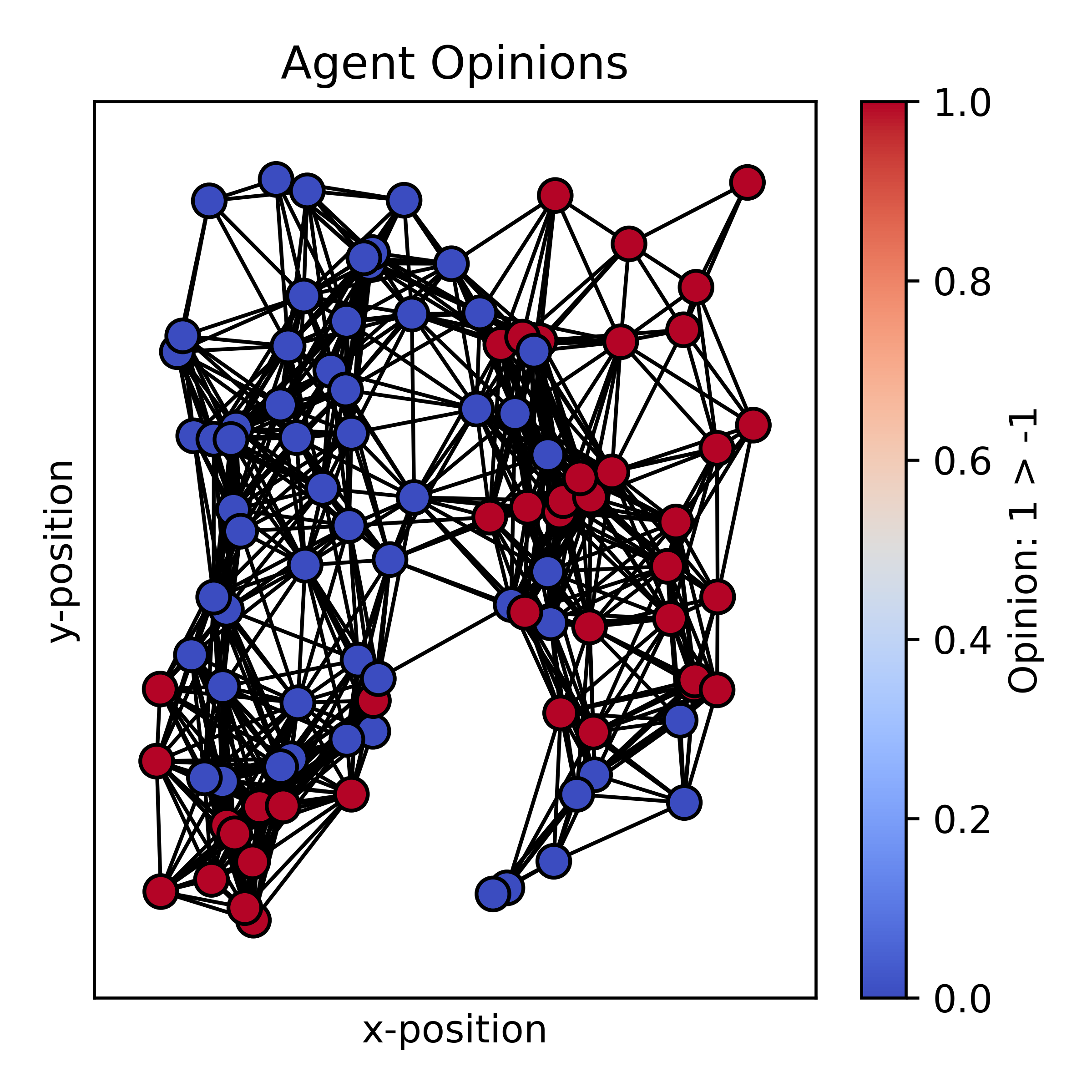}}\\
    \subfloat[]{\includegraphics[trim = {0 0 65 0}, clip = true, width = 0.27\textwidth]{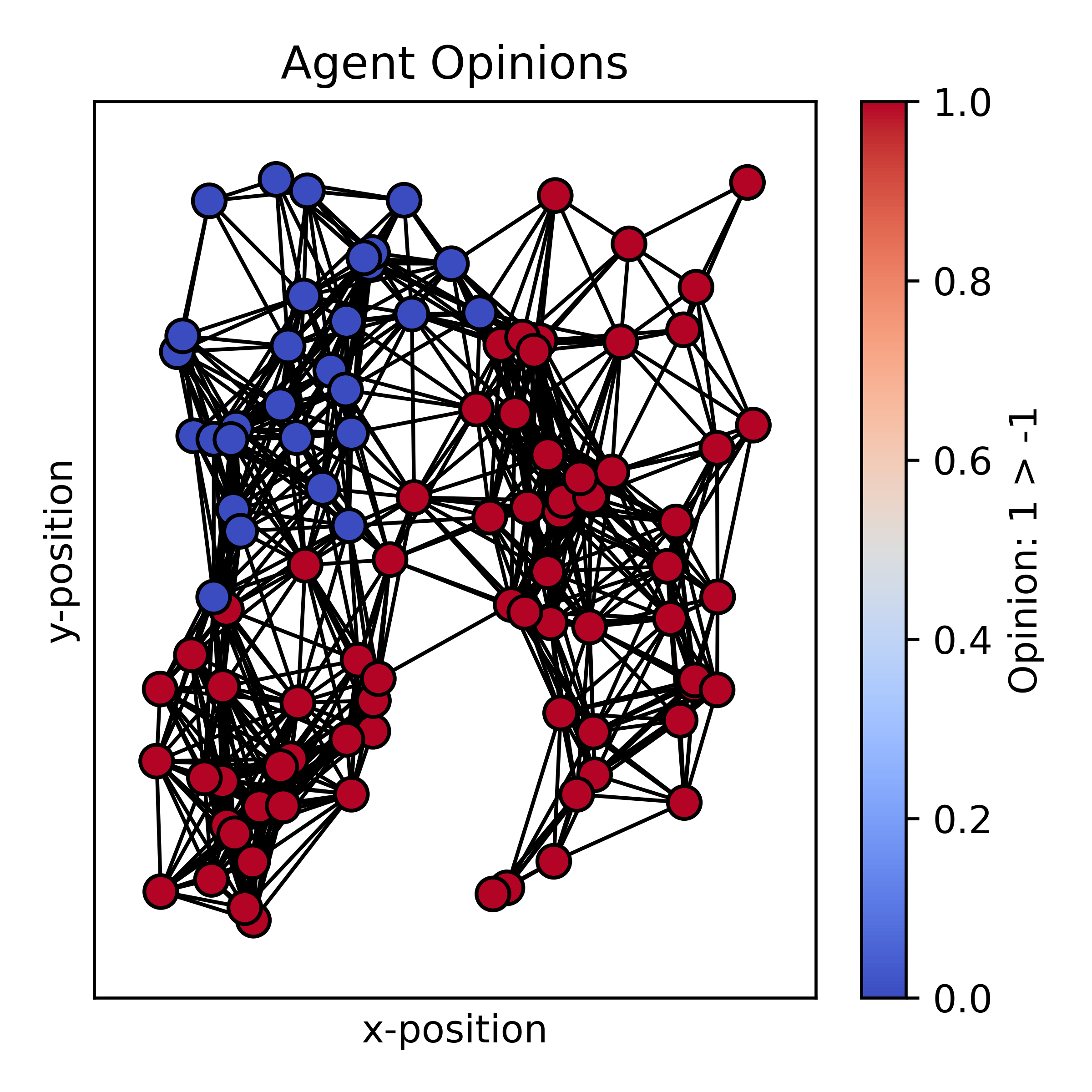}}
\subfloat[]{\includegraphics[trim = {0 0 65 0}, clip = true, width = 0.27\textwidth]{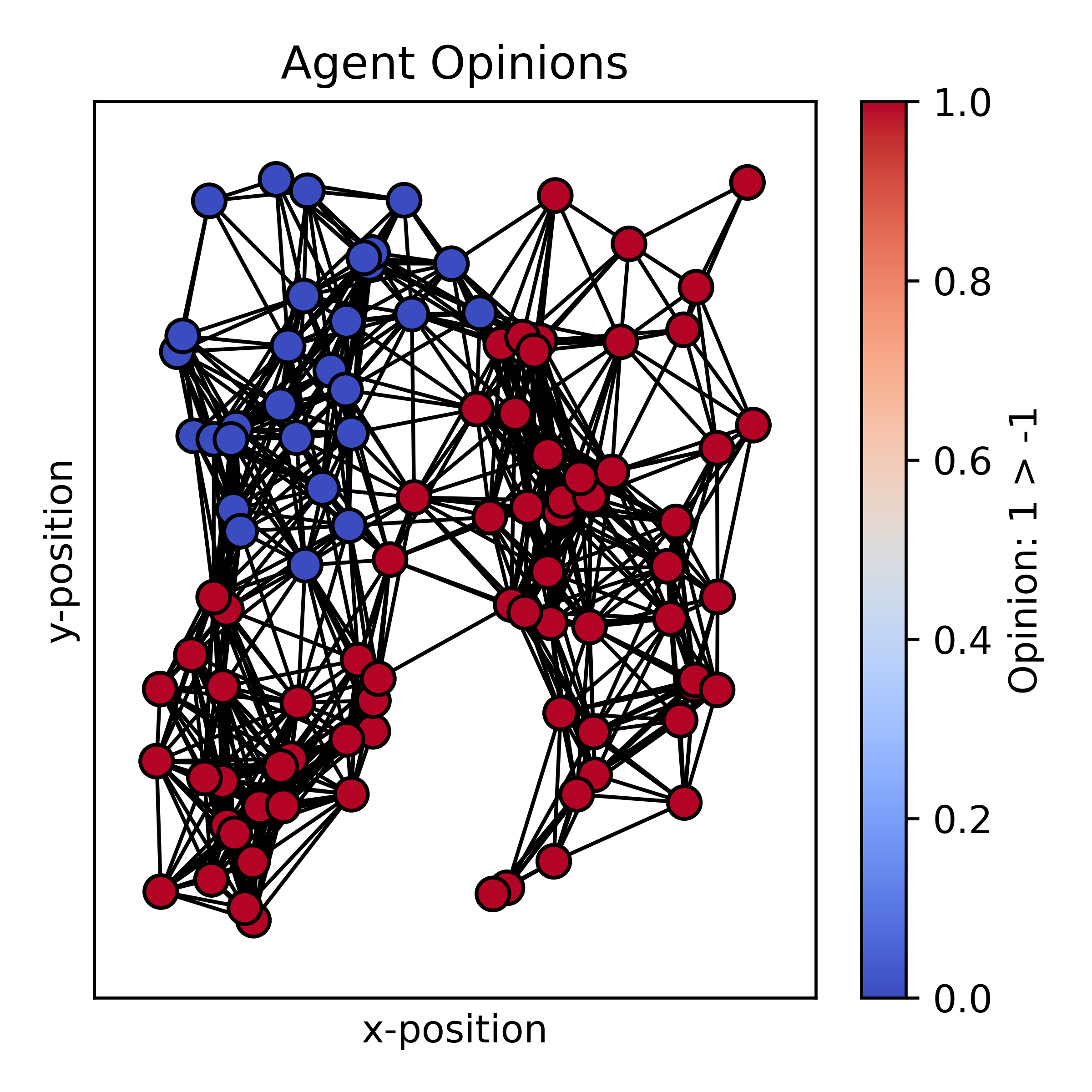}}
\subfloat[]{\includegraphics[trim = {0 0 65 0}, clip = true, width = 0.27\textwidth]{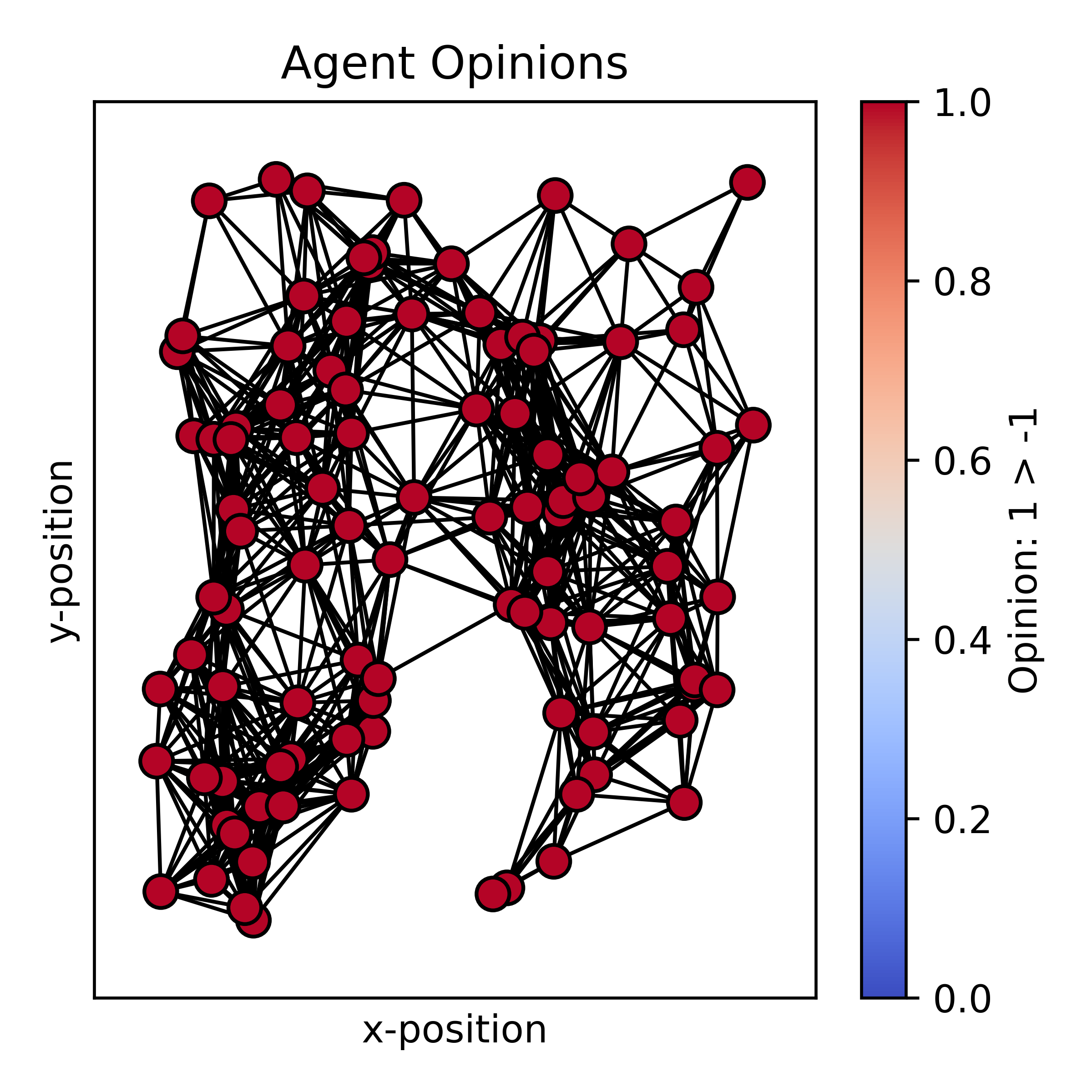}}
    \caption{Opinions in simulation run with metastable behaviour at timestamp {(A) $t=1$, (B) $t=100$, (C) $t=10^3$, (D) $t=10^4$, (E) $t=10^6$, and (F) $t\approx 5.82\times 10^6$.} Note the group with opinion $o=-1$ {(blue)} forms around $t=10^4$ {and switches to $o=1$ (red) after} $t=10^6$. The corresponding total number of agents holding opinion $o=1$ is plotted in Figure~\ref{fig:ms_total_log}. In this simulation run $r_g=0.25$, the other parameters are as in \S\ref{sec:sim_settings}.}
    \label{fig:ms_nets}
\end{figure}

To illustrate this phenomenon of metastability, we plot the state of the system at different points in time for a single trajectory. In Figure~\ref{fig:ms_total_log} we show the total number of agents holding opinion $o=1$ over time in this trajectory, which illustrates the metastable behaviour. In Figure~\ref{fig:ms_nets} we show the network of agents coloured according to their opinion at different time steps. {Note that because we select a run which illustrates metastable polarization, the network depicted here has more community structure than what may be typical of the random geometric network algorithm. This is because the presence of community structure allows nodes to have more in-community than out-community connections and so hold the opinion of their community for a long time (metastable), even if this is not uniform across communities.} We see that by $t=10^4$ two groups emerge; just less than 20 agents holding opinion $o=-1$ and the rest holding opinion $o=1$. This remains the case until shortly after time step $5.82 \times 10^6$ when the opinions all quickly converge to $o=1.$ The long time spent around one state with many small fluctuations followed by a quick exit to a stable state is typical of metastability. 

\subsection{Relationship to voter model}\label{sec:voter}
It is not clear from the simulations presented by BO or the simulations we have executed that consensus occurs with probability one. Indeed, polarisation may seem {persistent} because many simulation runs ended in a state of polarisation in both sets of numerical simulations. We know that consensus will be reached asymptotically, but how long the process may be in a state of polarisation is not addressed by Theorem~\ref{thm:con_g}.
To explore the stability of polarisation, we employ a separation of time scales argument which relates the ARLOD model to a different Markov chain, namely, the voter model.

It is well known~\cite{Phansalkar1994,Sastry1994,Borgers1997,Sato2002,Sato2003,Sato2005} that reinforcement learning dynamics can be described by the replicator dynamics in the continuous time limit, using a separation of times scales between agent learning and strategy adjustment. We now present a similar relationship between the ARLOD model and the jump chain (discrete time version) of the voter model~\cite{Holley1975} on a finite topology and in the case of two opinions. It is known that the voter model on scale-free networks~\cite{Suchecki2005a, Suchecki2005b}, and small-world networks~\cite{Vilone2004, Castellano2003} exhibits metastable polarisation and stable consensus.

\subsubsection{{Discrete time voter model}}\label{sec:method_voter}
In the voter model, nodes on a graph have an opinion, which may take one of two values ${-1,1}$. Repeatedly, a node is selected at random from the set of all nodes. This node performs an update in which it selects one of its neighbours and copies whichever opinion they have. Time may be indexed by each such round, or by a collection of rounds in which on average each node is selected once (on the order of the population size). {The version we discuss uses the former indexation of time.}

We define the discrete time voter model as a Markov chain $(X_t)_{t\geq 0}$ with $t\in \mathbb{N}$. As such, we define the graph on which the voter model is to take place $G = (V,E)$, with $V$ the set of vertices (voters) and $E$ the set of edges (connections between voters). The number of voters is $|V|=N$ and we endow each vertex $i$ with an opinion $o_i\in \{-1,1\}$ for $i\in \{1, \ldots, N\}$. As a result, the state space of the system is all possible assignments of each vertex to an opinion: $\mathcal{S}:= \{-1,1\}^N$.

We denote the unit vector of length $N$ with a one at the $l$-th entry and zeros everywhere else, as $\bm{e}_l$ for $l\in \{1,2,\ldots,N\}$.
The transition probability from state $\eta\in \mathcal{S}$ to state $\zeta \in \mathcal{S}$ is denoted $P_{\eta,\zeta} := \mathbb{P}(X_{t+1}=\zeta \mid X_t = \eta)$ and is given by
\begin{equation}\label{eq:P_voter}
    P_{\eta,\zeta} =\begin{cases}
    0 \quad &\text{if }\lVert \eta-\zeta \rVert_1 >2, \\
        \frac{1}{N}\left(\frac{1}{2}-\frac{o_l}{2d_l}\sum_{k\in N(l)}o_k\right) &\text{if }\lVert \eta-\zeta \rVert_1 =2, \text{and }\zeta=i - 2 o_{l}\bm{e}_l,\\
        1-\sum_{\zeta \neq \eta}P_{\eta,\zeta} &\text{if }\zeta=\eta. 
    \end{cases} 
\end{equation}
Here $d_l$ is the degree of voter $l\in V$ and $N(l)$ {$=\{u: (u,l)\in E\}$} is their neighbourhood in the graph $G$. 

Informally, the transition probability in (\ref{eq:P_voter}) is simply the uniform probability of agent $i\in \{1, \ldots, N\}$ being chosen, multiplied by the probability of them selecting a neighbour (uniformly at random) holding opinion $-o_j$. All transitions from $\eta\in \mathcal{S}$ to $\zeta\in \mathcal{S}$ in which the two states $\eta$ and $\zeta$ differ in more than one position occur with probability zero.

Given a starting assignment of opinions to voters $\eta\in \mathcal{S}$, the voter model is the Markov process $(X_t)_{t\geq0}$ that is Markov($\delta_\eta,P$), taking values in $\mathcal{S}$. Here $\delta_\eta$ is the delta function. Alternatively, given a distribution of the possible starting assignments of opinions to voters $\lambda$ such that $\mathbb{P}(X_0=\eta) = \lambda_\eta$ for each $\eta\in \mathcal{S}$, the voter model is Markov($\lambda,P$).

\begin{figure}
    \centering
    \includegraphics{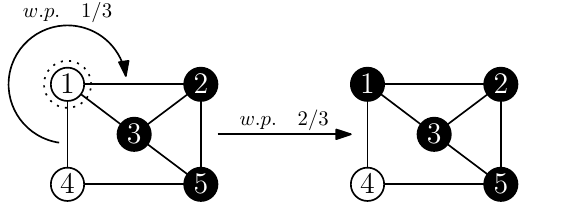}
    \caption{Illustration of the voter model dynamics. We show the transition probabilities conditioned on voter $1$ being selected to copy the opinion of one of their neighbours at random.}
    \label{fig:vm_ex}
\end{figure}

The dynamics of the voter model are illustrated in Figure~\ref{fig:vm_ex}. In this example, we consider $5$ voters, $V=\{1,\ldots,5\}$ with connections $E=\{(1,2),(1,3),(1,4),(2,3),(2,5),(3,5),(4,5)\}$ and initial opinions $X_0 = [-1,1,1,-1,1]$. We show the transitions conditioned on voter $1$ being selected to copy the opinion of one of their neighbours. In particular, if voter $1$ selects voters $2$ or $3$ they switch their opinion and if they select voter $4$ they keep their current opinion. These transitions occur with probability $2/3$ and $1/3$, respectively.

\subsubsection{{ARLOD model in batches}}\label{sec:method_batch}
The concept of multi-agent learning in batches has been explored in its own right~\cite{Barfuss2019,Barfuss2023,MeylahnJ2022}. It may be interpreted as a separation of time scales. That is, the rate at which agents learn about the behaviour of the environment or the other agents is faster than the rate at which they adjust their behaviour. Practically, it may be implemented by defining a batch size $b\in\mathbb{N}$ which constitutes a number of rounds in which the agent keeps their behaviour fixed and collects samples from their environment. At the end of this batch, the belief of the agent is updated using all the observations made during the batch.

Now we define the batch learning version of the ARLOD model. In particular, agent $i$ chosen to express their opinion in batch $t\in \mathbb{N}$ will express their opinion to their chosen neighbour $j$ in a batch of size $b_t\in\mathbb{N}$. 

That is, the dynamics follow the steps:
\begin{enumerate}
     \item At time $t$, an agent $i\in\{1,2,\ldots, N\}$ is selected uniformly at random from the population.
    \item This agent $i$ chooses a neighbour $j$ from their neighbourhood $N(i)$ uniformly at random.
    \item Then follow a sequence of subrounds indexed $s=1,\ldots,b_t$. Because agent $i$ is the only agent who can adjust their belief in this batch, we denote agent $i$'s Q-values in the subround $s$ by $Q'(s)$ and their opinion preference $Y_s'$ (with $Q'(0) = Q^i(t)$ and $Y_0' = Y^b_t(i)$). In each subround, agent $i$ expresses an opinion to agent $j$, following the rules of the ARLOD model:
    \begin{itemize}
        \item expressing their preferred opinion $o_{i}(s) = Y_s'$ at probability $1-\epsilon$,
        \item expressing their disfavoured opinion $o_{i}(s) = -Y_s'$ at probability $\epsilon$, and
        \item incorporating agent $j$'s honest response $R_j(s)$ into their $Q'$ value $Q_{o_{s}}'.$
    \end{itemize}
    Now we define the random batch size $b_t=\min\{s: Y_{s}' = Y_t(j)\}$, \textit{i.e.}, the number of subrounds required until agent $i$'s preference matches that of agent $j$.
    \item Agent $i$ updates their Q-values: $Q^i(t+1)\leftarrow Q'(t+b_t)$.
\end{enumerate}
We use this perhaps unconventional construction because the techniques in~\cite{Banisch2012} are not applicable here, as the states are not lumpable.

\begin{figure}[htb]
    \centering
    \includegraphics{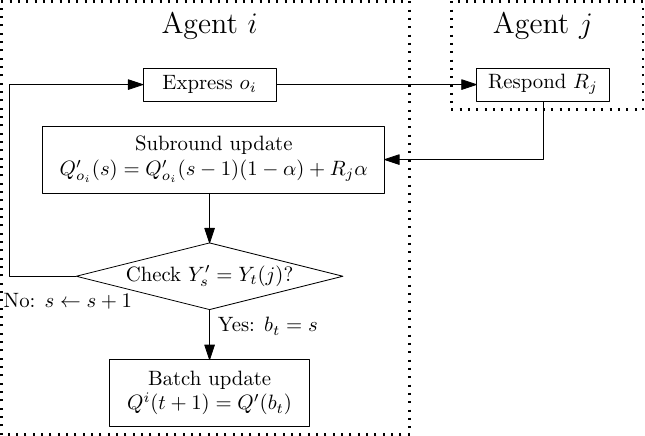}
    \caption{The dynamics in one time step of the batched version of the ARLOD model at a high level of abstraction. Agent $i$ expresses an opinion $b_t$ times to their neighbour agent $j$ who responds each time. Thereafter, agent $i$ updates their Q-values with all the feedback they received.}
    \label{fig:batch_Q}
\end{figure}

On a high level, the procedure of one such time step is depicted in Figure~\ref{fig:batch_Q}.

\subsubsection{{Relationship between the ARLOD model and the voter model}}

To establish the link between {the ARLOD model (with sophisticated agents) from computational sociology} and the voter model {(from sociophysics)}, we define the preference vector at time $t\in\mathbb{N}$: $Y_t$ whose elements are:
\begin{equation}
\label{eq:favouredopinions}
    Y_t(i) = \begin{cases}
        1 \quad &\text{if }Q_1^i(t) \geq Q_{-1}^i(t),\\
        -1 & \text{otherwise.}
    \end{cases}
\end{equation}
It takes values in the state space $\mathcal{S} := \{-1,1\}^N$.

Note that we use the weak inequality in (\ref{eq:favouredopinions}), though in the limit of interest, equality occurs with probability zero. We define the preference vector for the batched model as $(Y_t^b)_{t\geq 0}$. In essence, the batched model is a biased realisation of ARLOD; in the batch at time $t$ an agent is chosen to express their opinion to a neighbour as often ($b_t\in\mathbb{N}$ times) as is needed for them to have the same opinion preference. This occurs in the ARLOD model at probability $(1/N)^{b_t}$. For details, see \S\ref{sec:method_batch}.

We now state the main result of this section, which relates a batch learning version of the ARLOD model to the discrete time version of the voter model.
\begin{theorem}\label{thm:ql_voter_alt_batch}
    For any initial assignment of Q-values resulting in preference vector $Y_0 = \eta_0 \in \mathcal{S}$, the random process tracking the change of the preference vector $(Y^b_t)_{t\geq 0}$ in the batch version of the {ARLOD model} on graph $G$ converges in distribution to the voter model on the same graph:
    \begin{equation}
        \mathbb{P}(Y^b_t = \eta) = \mathbb{P}(X_t=\eta),\quad \forall \eta\in \mathcal{S},
    \end{equation}
    with $(X_t)_{t\geq 0}$ Markov($\delta_{\eta_0},P$) with $P$ as defined in (\ref{eq:P_voter}).
\end{theorem}

The proof is provided in Appendix~\ref{sec:proofvoter_alt_batch} and relies on the fact that an agent will receive enough feedback to make the ordering of their Q-values match that of their neighbour in finite time. Thus, we have shown that under a particular separation of time scales, the ARLOD model behaves like the discrete time voter model on a finite graph with two opinions. The construction of the batched ARLOD model and its relation to the voter model ensures that any state that is metastable in the voter model will also be metastable in the ARLOD model. {This is because any realisation of events in the batched ARLOD model also occur with positive probability in the standard ARLOD model.}

\subsection{Instability of consensus and ergodicity of symmetric reinforcement learning}
We now introduce a new model based closely on the ARLOD model, with a subtle difference: both agents involved in an interaction express their opinion in the same way and update their Q-values as a result of what they observe. Because now the roles of the two agents are indistinguishable, we call this the \textit{symmetric} reinforcement learning for opinion dynamics (SRLOD) model.

\subsubsection{{The symmetric reinforcement learning opinion dynamics (SRLOD) model}}\label{sec:method_sym}
A population of $n\in\mathbb{N}$ agents is embedded in a random (connected) geometric network topology. In each discrete time step $t\in\mathbb{N}_{\geq 0}$ an edge $(i,j)\in E$ is selected uniformly at random. The two agents on either end of this edge $i$ and $j$ express an opinion to one another $o_i(t), o_j(t)\in\{-1,1\}$. Subsequently, both agents update the Q-value of their expressed opinion as follows:
\begin{align}
    q_{o_i(t)}^i(t+1)& =q_{o_i(t)}^i(t)(1-\alpha) + \alpha o_i(t) o_j(t) \\
    q_{o_j(t)}^j(t+1)& =q_{o_j(t)}^j(t)(1-\alpha) + \alpha o_i(t) o_j(t),
\end{align}
where $\alpha\in(0,1)$ is the learning rate. To differentiate it from the ARLOD model, we let $q_o^i(t)$ denote the Q-value agent $i\in \{1, \ldots, N\}$ has for opinion $o\in \{-1,1\}$ at time $t\in\mathbb{N}$. The Q-value of the opinion the agents did not express is not updated. We call the opinion $o$ such that $q_o^i(t)>q_{-o}^i(t)$ agent $i$'s preferred opinion. We assume that both agents express their preferred opinion with probability $1-\epsilon$ (called exploitation) and express their disfavoured opinion with probability $\epsilon$ (called exploration). 

The difference thus between this model and the original model is only that instead of a one-sided interaction, both agents may explore and learn from the interaction each round. 

\subsubsection{{Instability of consensus in the SRLOD model}}
We show that consensus is no longer stable in this model.  

\begin{lemma}[Consensus is not stable]\label{lem:con_not}
    If there exists a time $t_0\in\mathbb{N}$ such that $q_o^i(t_0)>q_{-o}^i(t_0)$ for some opinion $o\in \{-1,1\}$ and each agent $i\in \{1,2,\ldots,N\}$, then 
    \begin{equation}
        \mathbb{P}(q_o^i(t)>q_{-o}^i(t), \forall t\geq t_0)=0.
    \end{equation}
\end{lemma}
The proof of Lemma~\ref{lem:con_not} is presented in Appendix~\ref{sec:proof_con_not}. This and the next result depend on the fact that any sequence of actions has positive probability in this model because \emph{both} agents learn from an interaction and explore with probability $\epsilon>0$. In particular, the probability of any finite sequence of actions of length $l<\infty$ occurs with a probability bounded from below by $p(l)$:
    \begin{equation}\label{eq:path_l}
       p(l)= \left(\frac{1}{|E|}\right)^l (\epsilon^2)^l>0.
    \end{equation}

Consensus not being a stable state is a fundamental difference between the symmetric and the asymmetric model. To elucidate this difference, we introduce the preference vector, $y_t$, of length $N$, whose $i$-th element takes the value:
\begin{equation}
    y_t(i) = \begin{cases}
        1 \quad &\text{if } q_1^i(t) \geq q_{-1}^i(t) \\
        -1 &\text{otherwise.}
    \end{cases}
\end{equation}
The preference vector describes which opinion ($1$ or $-1$) each agent $i\in\{1,2,\ldots,N\}$ favours.
The dynamics of the preference vector are ergodic in the symmetric model. 
\begin{proposition}[Time-evolution of the preference vector is ergodic]\label{prop:erg}
    The probability of the preference vector transitioning in finite time between any two states $\eta,\zeta\in \mathcal{S}$ is positive, i.e.,
    \begin{equation}
        \mathbb{P}(y_{t_0+k}=\zeta\mid y_{t_0}=\eta)>0 ,\quad \text{for all }\eta,\zeta\in\mathcal{S}, \text{and some } k<\infty.
    \end{equation}
\end{proposition}
To prove this, one first delineates a finite sequence between any two states $\eta,\zeta\in\mathcal{S}$ and observes that the probability of these sequences is positive by (\ref{eq:path_l}). For any two states there is a finite sequence of events which leads from one to the other as all agents take both actions with positive probability, and can always switch their belief in a finite number of rounds. Thus, all states communicate with one another. The ergodicity of the SRLOD model is illustrated in Appendix~\ref{app:SRLOD}. {The SRLOD model being ergodic means that looking for the probability of consensus, or polarization is no longer as straightforward because both of these occur with probability one over the infinite time horizon. Instead, it is reasonable, for example, to calculate or estimate the stationary distribution of the model which gives insights into the relative time spent in consensus and polarized states. This illustrates how the tools used to study a model differ based on the seemingly innocuous assumption of asymmetry in the agent-to-agent interaction.}


\section{Discussion}\label{sec:conc} 
We have analysed the ARLOD model of social learning put forth by Banisch and Olbrich~\cite{Banisch2019}. Our first main theorem shows that consensus is reached asymptotically with probability one for any finite {and connected} population structure. In particular, this is in contrast with the {persistence} of polarisation originally reported for that model. A small modification of that model, based on symmetrizing the interaction-learning relation between the agents, results instead in ergodic dynamics, which thus destabilizes consensus somewhat. This result mirrors the difference between the voter model and the \emph{noisy} voter model, in which a random probability of switching one's opinion is introduced \cite{granovsky1995noisy, carro2016noisy}. 

The highlighted importance of network structure in the original article~\cite{Banisch2019} warrants attention. The theoretical arguments we present here to show that a) consensus is the only stable state in the original model and b) that the symmetrized model is ergodic required only that the network is connected. 
Thus, \textit{qualitatively}, the assumption of network structure is not very important. We do, however, see that it plays a{n important} role \textit{quantitatively} in the time taken for consensus to be reached. {It is likely that the metastability of polarization emerges because of strong community structure in the network. This is in accordance with previous findings of the effect of network structure on the timescales of the resulting dynamics (see for instance~\cite{Centola2015,Zarei2024})} Many studies on polarisation and other social dynamics focus on the importance of network structure. {In addition to investigating the effects of networks, it may also be} important to disentangle which outcomes of the model are truly caused by networks structure and which outcomes are {the result of} other --- more implicit --- modelling decisions, {such as asymmetry in the agent-to-agent interaction}.

Having proved that the polarisation observed in the ARLOD model is not stable and that consensus is guaranteed, we turn to the original research question. What causes stable polarisation? We provide conditions (a systematic biasing) for which the ARLOD model converges to the voter model. Polarisation can be metastable in the voter model, and, by their relation, also in the reinforcement learning model. This {(relationship between the ARLOD model and the voter model)} bridges multiagent learning models and models well studied in sociophysics and theoretical biology. 
 
Our results raise questions regarding the possibility of finding a model of opinion dynamics excluding repulsive forces and allowing for stable polarisation. Can we say that a reasonable model of opinion dynamics should exhibit stable {(as defined in Definition~\ref{def:stable})} polarisation? Is the polarisation we observe around us stable or metastable? Future research is required to give an example of such a model or a proof that it does not exist. These questions might be explored by investigating learning in the `real world' (to identify appropriate $\alpha$ and $\epsilon$) as well as the influence of parameter values $\alpha$ and $\epsilon$ in the ARLOD model. It could be that `real world' learning is such that consensus would be reached quickly under the ARLOD model, indicating that a more realistic model requires additional elements. Alternatively, it may be that the parameters of the `real world' are such that the time it takes to exit the metastable polarised state is so long that differentiating between metastable and stable polarisation in the real world is difficult.

{A limitation of the models we study is that the memory of the agents is entirely implicit because we use stateless Q-learning. Explicit inclusion of memory may be done by using Q-learning \textit{with} states, where each state corresponds to the last action taken by each of an agents neighbours. This would complicate the analysis: If additionally to memory, an agent knows the identity of the neighbour they are expressing their opinion to, it is possible that polarization becomes absorbing (and therewith stable). Note that care would have to be taken to determine how an agent responds to an opinion to avoid decoupling the dynamics of each pair of agents from other interactions. Another limitation of the model relates to the isolation of the dynamics from other influences. Effects other than social influence that may be driving agent opinions are an internal cognitive process related to their opinion such as in~\cite{Searle2023,Giardini2015}, or pressure applied by mass media to follow a certain opinion~\cite{Tornberg2022,Hoffman2007,Pansanella2023,VanSanten2024}. Finally, the assumption that the agents of the model are fixed (no new agents enter, or old ones leave) can be seen as unrealistic. Important to note that changing this assumption may change the outcome of the analysis. In particular if new agents have random Q-values, this destabilizes consensus in the ARLOD model.} 


\section{Methods}
\subsection{ARLOD simulation settings}\label{sec:sim_settings}

We have chosen the parameter settings based on the following considerations. A greater number of agents means that more rounds are required to select each agent sufficiently often to reach consensus. On the other hand, a smaller learning rate increases what `sufficiently often' means per agent, as indicated in (\ref{eq:r_switch}). To strike a balance between these effects, we set $N=100$ and $\alpha=0.25$. Following BO, we set $\epsilon = 0.1$ and initialise the Q-values uniformly in $[-0.5,0.5]$. The radius for the random geometric graph model $r_g\in[0.25,0.5]$ is selected to exhibit a range of behaviour, focusing on connected graphs. We have chosen the maximum time to simulate ($10\times 10^9$ rounds) and the number of simulation iterations (500 iterations) to be significantly greater than those used by BO ($2\times 10^6$ rounds and 100 iterations). This allows the simulation to reach consensus more frequently, which we know occurs eventually with probability one (by Theorem \ref{thm:con_g}).

\subsection{Random connected geometric graphs}\label{sec:method_graph}
The algorithm to generate a connected random graph is provided in Appendix~\ref{app:geograph}. We use the subroutine for the generation of a random geometric network from the Python NetworkX package~\cite{NetX}. For a detailed discussion on random geometric graphs and their properties, the interested reader is referred to~\cite{Dall2002,Penrose2003}. The random geometric graph model is popular in the context of social dynamics because it mimics the homophily of real social networks {as claimed by}~\cite{McPherson2001}.

The general idea of the random geometric graph is to distribute the desired number of nodes randomly in Euclidean space (we use $[0,1]^2$) and fixes a radius $r_g$. Subsequently, any nodes $u,v$ that are distance $d(u,v)<r_g$ from one another are connected by an edge $(u,v)$. Because we are interested in connected networks, we simply repeat the standard procedure until a connected graph is sampled. We take care to only use $r_g $ for which the probability of sampling a connected graph is sufficiently high (as described in \S\ref{sec:sim_settings}).

\bibliographystyle{ieeetr}
\bibliography{Biblio.bib}

\appendix

\section{Proof of Lemma~\ref{lem:stay}}\label{sec:proofstay}
\begin{proof}
We proceed by induction on time. Suppose that $Q_o^i(t_0)>Q_{-o}^i(t_0)$ at time $t_0\geq 0$, for some opinion $o\in\{-1,1\}$, and each agent $i\in \{1,2,\ldots,N\}$.

The base case is that in round $t_0+1$, the ordering of all the Q-values will remain the same.

In round $t_0$, any agent $i\in\{1,2,\ldots, N\}$ may be chosen to express their opinion to one of their neighbours.

\textbf{Case 1.} Suppose they exploit their preferred opinion (the one with greater Q-value). Any agent they express their opinion to, has the same ordering among their Q-values by the conditions of the lemma, and so responds with an action that leads to a positive reward. Thus,
\begin{equation}
    Q_o^i(t_0+1) \geq Q_o^i(t_0),
\end{equation}
the Q-value of the preferred opinion in round $t_0+1$ is at least as great as in $t_0.$

\textbf{Case 2.} Suppose they explore by taking the action with lesser Q-value. Any neighbour they express this opinion to responds honestly. By the assumption, all agents have the same Q-value ordering, so the honest response to exploration is an action that leads to a punishment. Thus, 
\begin{equation}
    Q_{-1}^i(t_0+1)\leq Q_{-1}^i(t_0),
\end{equation}
the Q-value of the disfavoured action in round $t_0+1$ is lower than or equal to what it was in round $t_0$. This is true because the Q-values are initialised to be in $[-1,1]$ and will stay therein indefinitely by the updating prescribed.

This proves the base case (as this holds for all agents that could have been chosen in round $t_0$): $Q_o^i(t_0+1)>Q_{-o}^i(t_0+1)$ for all agents $i\in \{1,2,\ldots,N\}$. 

In the induction step we assume it is true until rounds $t_0+n$ for $n>0.$ To show that it is true for all rounds up until $t_0+n+1$, we simply follow the same procedure as in the base case but for the game in round $t_0+n$ which determines the Q-values in round $t_0+n+1$.
\end{proof}

\section{Proof of Lemma~\ref{lem:reach}}\label{sec:proofreach}
\begin{proof}
    First, we delineate a sequence of events of finite length which may lead from any state to consensus. Secondly, we will show that this sequence of events has positive probability. 

    Suppose agent $i$ favours opinion $o$ and has a neighbour $j$ who prefers opinion $-o$, all at time $t_{0}$. If agent $i$ is drawn to express their opinion to agent $j$ every round for $L\in\mathbb{N}$ rounds and always exploits their preferred opinion, the Q-value for this opinion is given by:
    \begin{equation}
        Q_{o}^i(t_{0}+l) = Q_{o}^i(t_{0}+l-1)(1-\alpha) -\alpha,
    \end{equation}
    for all $l=1,2,\ldots, L$. A term by term comparison shows that this is bounded from above by
    \begin{equation}
        Q_{o}^i(t_{0}+l) \leq Q_{o}^i(t_{0}+l-1)(1-\alpha),
    \end{equation}
    since $\alpha \in (0, 1)$. Thus, an upper bound of the Q-value in round $t_{0}+l$ is given by $Q_{o}^i(t_{0})(1-\alpha)^l$ for all $l = 1,2,\ldots, L$ as long as $Q_o^i(t_0)>0$\footnote{When both Q-values have the same sign, only one of them needs to be adjusted in the way described here until it changes sign.}.

    Subsequently, if agent $i$ is drawn to express their opinion to agent $j$ another $M\in\mathbb{N}$ times and explores their disfavoured action in each of these rounds, this opinion's Q-value follows:
    \begin{equation}
        Q_{-o}^i(t_0+L+m) = Q_{-o}^i(t_0+L+m-1)(1-\alpha) +\alpha,
    \end{equation}
    for all $m=1,2,\ldots,M$. A term by term comparison shows that this is bounded from below by
    \begin{equation}
        Q_{-o}^i(t_0+L+m)\geq Q_{-o}^i(t_0+L+m-1)(1-\alpha).
    \end{equation}
    Again a lower bound to this Q-value in round $t+0+L+m$ is given by $Q_{-o}^i(t_0+L)(1-\alpha)^m$ as long as $Q_{-o}^i(t_0+L)<0$.

    We bound from above the number of rounds needed for any agent's opinion to be switched, by the number of rounds needed should they start as far away from one another as possible, $\bm{Q} = (-1,1), $ or $(1,-1)$ and be set to cross at zero. The Q-value of the originally preferred opinion reaches $\xi\in(0,\alpha)$ at least by the lowest integer $r$ which satisfies:
    \begin{equation}
        Q_o^i(t_0)(1-\alpha)^{r}\leq \epsilon,
    \end{equation}
    if they express this preferred opinion in each round. Dividing by $Q_o^i(t_0)$, taking the logarithm on both sides and rearranging we get
    \begin{equation}
    r = \bigg\lceil \frac{\log (\xi/Q_o^i(t_0))}{\log  (1-\alpha)}\bigg\rceil.
    \end{equation}
    By a similar procedure we see that the Q-value of the originally disfavoured opinion reaches $-\epsilon$ after a further $r$ interactions (of exploring in each subsequent round). After two more interactions in which the agent expresses each opinion once, the Q-value ordering has switched:
    \begin{equation}
        Q_{-o}^i(t_0+2r+2)\geq -\xi (1-\alpha) + \alpha > \xi (1-\alpha) -\alpha \geq Q_o^i(t_0+2r+2), 
    \end{equation}
    as long as $\alpha>\xi$, which is satisfied by an appropriate choice of $\xi$. 
    
    The number of rounds this takes is $2r+2$. The probability that this happens is bounded by the probability of the agent $j$ being drawn to express their opinion to agent $k$ $2r+2$ times, multiplied by the probability that they take the required action in each round. This is a lower bound because it does not matter whether agent $i$ first exploits $r+1$ times and then explores $r+1$ times in that order. It only matters that there is a total of $r+1$ explorations and exploitations in the $2r+2$ rounds. Thus, the probability $p_{\text{switch}}$ of one agent switching their opinion (if they have at least one neighbour that disagrees with them) is lower bounded by
\begin{equation}
    p_{\text{switch}} > (1-\epsilon)^{r+1} \epsilon^{r+1} \left(\frac{1}{N(N-1)}\right)^{2r+2}>0.
\end{equation}
    Here, agent $i$ is drawn to express their opinion with probability $1/N$ and we bound the probability that they express this opinion to agent $k$ from below by $1/(N-1)$ as that is the maximum possible degree for any agent in the network. This probability is greater than zero simply because it is a finite product of positive numbers.

    In a connected population of $N$ agents which is not yet in consensus, there is always at least one edge which has an agent who prefers opinion $o$ on one side and opinion $-o$ on the other side. Furthermore, in the initial state there are at most $N-1$ agents who prefer the `wrong' opinion at time $t_0$. So with probability $p>p_{\text{switch}}^{N-1}>0$, in $(N-1)(2r+2)<\infty$ rounds all $N$ agents hold the same opinion.
\end{proof}

\section{Proof of Theorem~\ref{thm:ql_voter_alt_batch}}\label{sec:proofvoter_alt_batch}

\begin{proof}
    At time $t=0$, by definition, we have that $\mathbb{P}(Y_0=\eta_0)=1$ which is also true for $X_t$ under $\delta_{\eta_{0}}$: $\mathbb{P}(X_0=\eta_{0}) =1.$

    Next we show that for $t\in\mathbb{N}$,
    \begin{equation}
      \mathbb{P}(Y_{t+1}= \eta\mid Y_t = \eta_t,\ldots,Y_0=\eta_0) = \mathbb{P}(Y_{t+1}= \eta\mid Y_t = \eta_t) = P_{\eta_t,\eta}.
    \end{equation}
The $Y_{t+1}$'s independence on $Y_{t-1},\ldots,Y_0$ follows from the fact that in the batch at time $t\in\mathbb{N}$ the agents determine their behaviour entirely from the state $Y_t$. Expressing agents express their favoured opinion with probability $1-\epsilon$ and express their disfavoured opinion with probability $\epsilon.$ Responding agents always do so honestly, rewarding their favoured opinion and punishing their disfavoured opinion.

Note that $\mathbb{P}(Y_{t+1}=\eta\mid Y_t=\eta_t)=0$ whenever $\lVert \eta-\eta_t\rVert_1>2$, just as in (\ref{eq:P_voter}). This is because as soon as $\lVert \eta-\eta_t\rVert_1>2$ we have that more than one agent has switched their opinion after the batch at time $t$. This is impossible because only one agent updates their Q-values during a batch.

We proceed in two cases, one when $\lVert \eta-\eta_t\rVert_1=2$ and the other when $\lVert \eta-\eta_t\rVert_1=0$. Note that $\lVert \eta-\eta_t\rVert_1 \neq 1$ because $\eta,\eta_t\in \mathcal{S} = \{-1,1\}^N$.

\textbf{Case 1.} $\lVert \eta-\eta_t\rVert_1=2$: In this case states $\eta$ and $\eta_t$ differ in exactly one position which, without loss of generality, we label $l$. In order for agent $l$ to switch their favoured opinion in batch $t$, they must be selected uniformly at random to express their opinion to one of their neighbours. This happens at probability $1/N$.

If an agent expresses their opinion to an agent who favours opinion $-o_l$, agent $l$, gets punished for each time they express their favoured opinion and rewarded for each time they express their disfavoured opinion. Thus, if agent $l$ expresses their favoured opinion at least $r+1$ times and their disfavoured opinion at least $r+1$ times, where
\begin{equation}
     r = \bigg\lceil \frac{\log (\epsilon)}{\log  (1-\alpha)}\bigg\rceil,
\end{equation}
then their opinion will have switched (see the proof of Lemma~\ref{lem:reach}). 
Thus, we can lower bound the probability of the agent switching their opinion after $2r+2$ rounds by 
\begin{equation}
    p_{\text{switch}}\geq \epsilon^{r+1}(1-\epsilon)^{r+1}.
\end{equation}
Then the probability that the agent does not switch their opinion in finite time is upper bounded by $\lim_{k\to\infty}(1-p_{\text{switch}})^k = 0$. This means that if the agent has selected to express their opinion to an agent who favours opinion $-o_l$, they will switch their opinion, and this will happen in finite time.

The probability of agent $l$ switching their opinion is given by the probability that they select a neighbour favouring the opposite opinion to theirs. As before, we denote $d_l$ as the degree of agent~$l$. Furthermore, we denote with $a(l)$ and $c(l)$ the number of agents in $l$'s neighbourhood who are in agreement and contradiction with $l$ respectively. Then, because agents select a neighbour uniformly at random, the probability of $l$ switching their opinion is
\begin{equation}
    \mathbb{P}(o_l\to -o_l) = \frac{c(l)}{d_l}.
\end{equation}
This may be rewritten and rearranged as follows:
\begin{align}
    \mathbb{P}(o_l\to -o_l) &= \frac{c(l)}{2d_l} + \frac{c(l)}{2d_l}\\
    &=  \frac{c(l)}{2d_l} + \frac{d_l-a(l)}{2d_l}\\
    &= \frac{1}{2}\left( 1+ \frac{c(l)-a(l)}{d_l}\right).
\end{align}
To extract this from $Y_t$, notice that it can be rewritten as
\begin{equation}
    \mathbb{P}(o_l\to -o_l) = \frac{1}{2}\left(1-\frac{Y_t(l)}{d_l}\sum_{k\in N(l)}Y_t(k)\right),
\end{equation}
where $N(l)$ is the neighbourhood of agent~$l$ in the graph $G$. Multiplying this by $1/N$, the probability that agent~$l$ is selected to express their opinion in the first place, we get precisely $P_{\eta_t,\eta}$ as required.

\textbf{Case 2.} $\lVert \eta-\eta_t\rVert_1=0$: In this case, an agent is selected to express their opinion, and they do so to a neighbour who is in agreement with them (in which case $b_t=1$). Given that agent~$l$ is selected, this happens with probability 
\begin{equation}
    \mathbb{P}(o_l\to o_l) = 1-\frac{1}{2}\left(1-\frac{Y_t(l)}{d_l}\sum_{k\in N(l)}Y_t(k)\right).
\end{equation}
This is exactly $1-N P_{\eta_t,\eta}$ for $\eta = \eta_t-2Y_t(l)\bm{e_l}$. Summing over all agents that might be selected and multiplying by the probability of selecting those agents, we get the required probability of transitioning to the same state,
$P_{\eta_t,\eta_t}=1-\sum_{\eta\neq \eta_t}P_{\eta_t,\eta}$.
\end{proof}

\section{Proof of Lemma~\ref{lem:con_not}}\label{sec:proof_con_not}
\begin{proof}
    We show that a sequence of events which leads from consensus to not-consensus is of finite length and positive probability.

    Observe that each agent explores (expresses and reinforces the disfavoured opinion) at probability $\epsilon>0$. Observe also that if this disfavoured opinion is reinforced maximally $\kappa +1$ times with 
    \begin{equation}
        \kappa =  \bigg\lceil \frac{\log (\epsilon)}{\log  (1-\alpha)}\bigg\rceil,
    \end{equation}
    which is finite, and similarly if their preferred opinion is punished $\kappa +1$ times, that they switch ordering of opinion Q-values \footnote{In similar fashion to as it was shown in the proof of Lemma~\ref{lem:reach} for the asymmetric model.}.

    Given a sequence of agent actions, the probability that they take the action in some round required by the sequence, is always bounded from below by $\epsilon>0$. This is because they express their disfavoured opinion at probability $\epsilon$ and their favoured opinion at probability $1-\epsilon>\epsilon$.

    This means that the probability of any finite sequence of actions of length $l<\infty$ occurs at a probability bounded from below by $p(l)$ defined in (\ref{eq:path_l}).
    Thus, the probability of an agent exploring and being rewarded $\kappa+1$ times and exploiting and being punished $\kappa +1$ times is positive because this is a sequence of events with length $2\kappa+2$. This is the maximal length sequence which leads to one agent changing the ordering of their Q-values. So we have that the probability that such a switch \textit{never} happens (the probability of consensus for all $t\geq t_0$):
    \begin{equation}
        \mathbb{P}(q_o^i(t)>q_{-o}^i(t), \forall t\geq t_0) \leq \lim_{k\to \infty}(1-p(2\kappa+2))^k = 0.
    \end{equation}
    Therefore consensus is not absorbing (and not a stable state). 
\end{proof}

\section{Algorithm to generate a geometric random graph}\label{app:geograph}
\begin{algorithm}[H]
\SetAlgoLined
\SetKwInOut{Input}{Input}
\SetKwInOut{Output}{Output}

\Input{$N$: number of nodes, $r_g$: radius}
\Output{$G$: connected random geometric graph}

\BlankLine
$Check \leftarrow$ True\;
\While{$Check$}{
$G \leftarrow$ empty graph\;
\For{$i \leftarrow 1$ \KwTo $N$}{
    Add node $i$ with random coordinates in $[0,1]^2$ to $G$\;
}
\For{each pair of nodes $(u, v)$ in $G$}{
    \If{distance($u$, $v$) $\leq r_g$}{
        Add edge $(u, v)$ to $G$\;
    }
}
\If{$G$ is connected}{
$Check \leftarrow$ False
}
}
\caption{Generate connected random geometric graph with $N$ nodes and radius $r_g.$}
\end{algorithm}

\section{Illustration of the ergodicity in the SRLOD model}
\label{app:SRLOD}
\begin{figure}[htb]
    \centering
7    \includegraphics[width =0.99\textwidth]{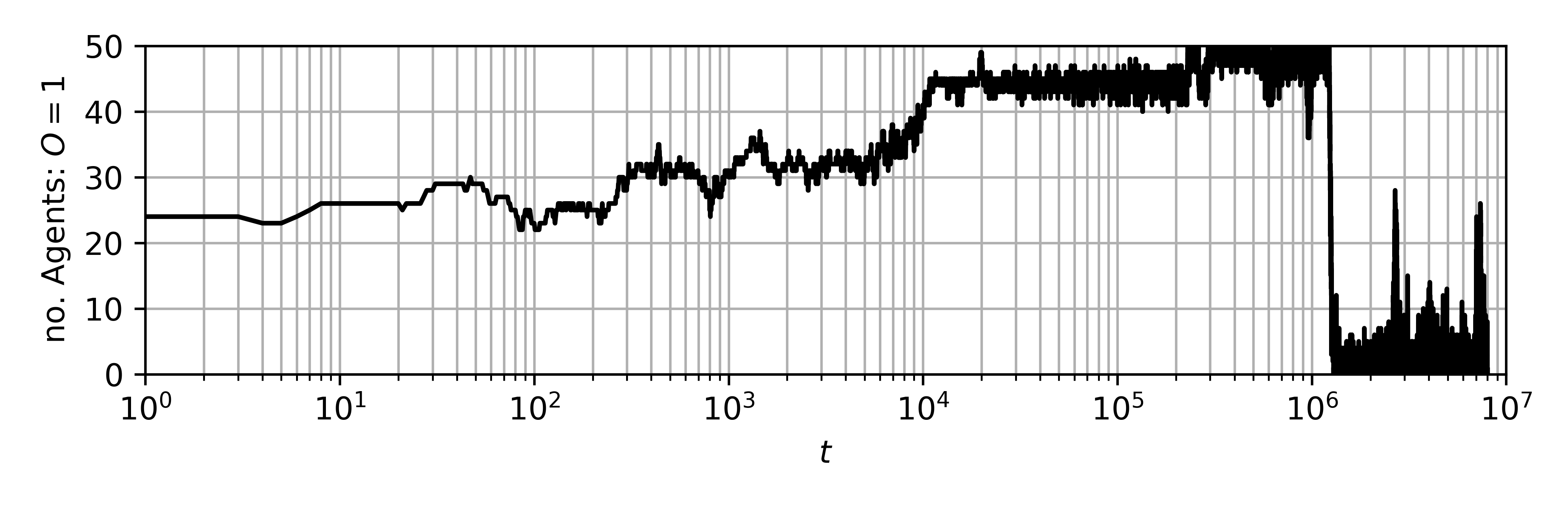}
    \caption{The number of agents holding opinion $o=1$ in a simulation run of the SRLOD model plotted with time on a logarithmic scale. Notice the switch from consensus on opinion $o=1$ to opinion $o=-1$ shortly after $t=10^6$.}
    \label{fig:erg_total_log}
\end{figure}

\begin{figure}[htb]
    \centering
    \subfloat[]{\includegraphics[trim = {0 0 65 0}, clip = true, width = 0.27\textwidth]{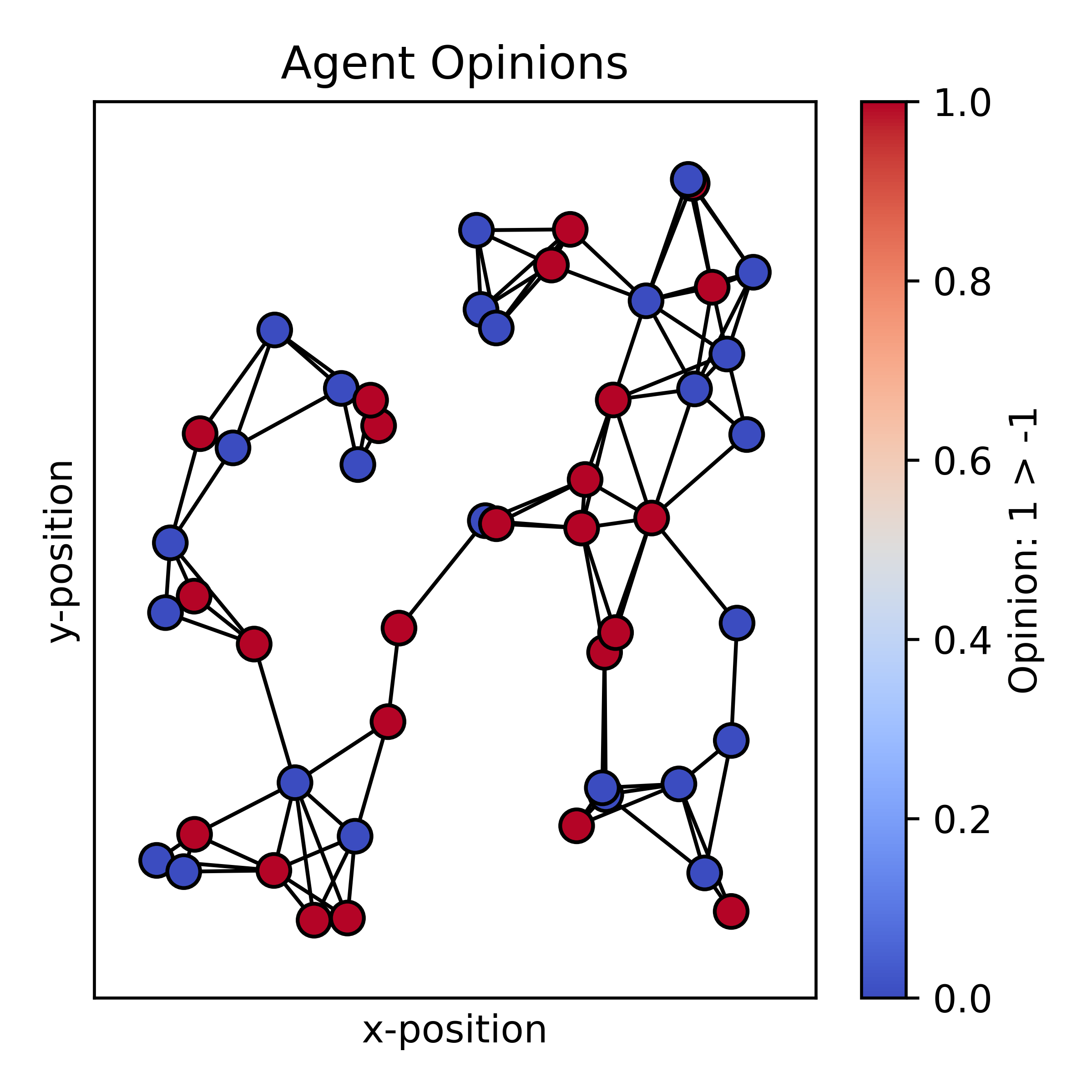}}
\subfloat[]{\includegraphics[trim = {0 0 65 0}, clip = true, width = 0.27\textwidth]{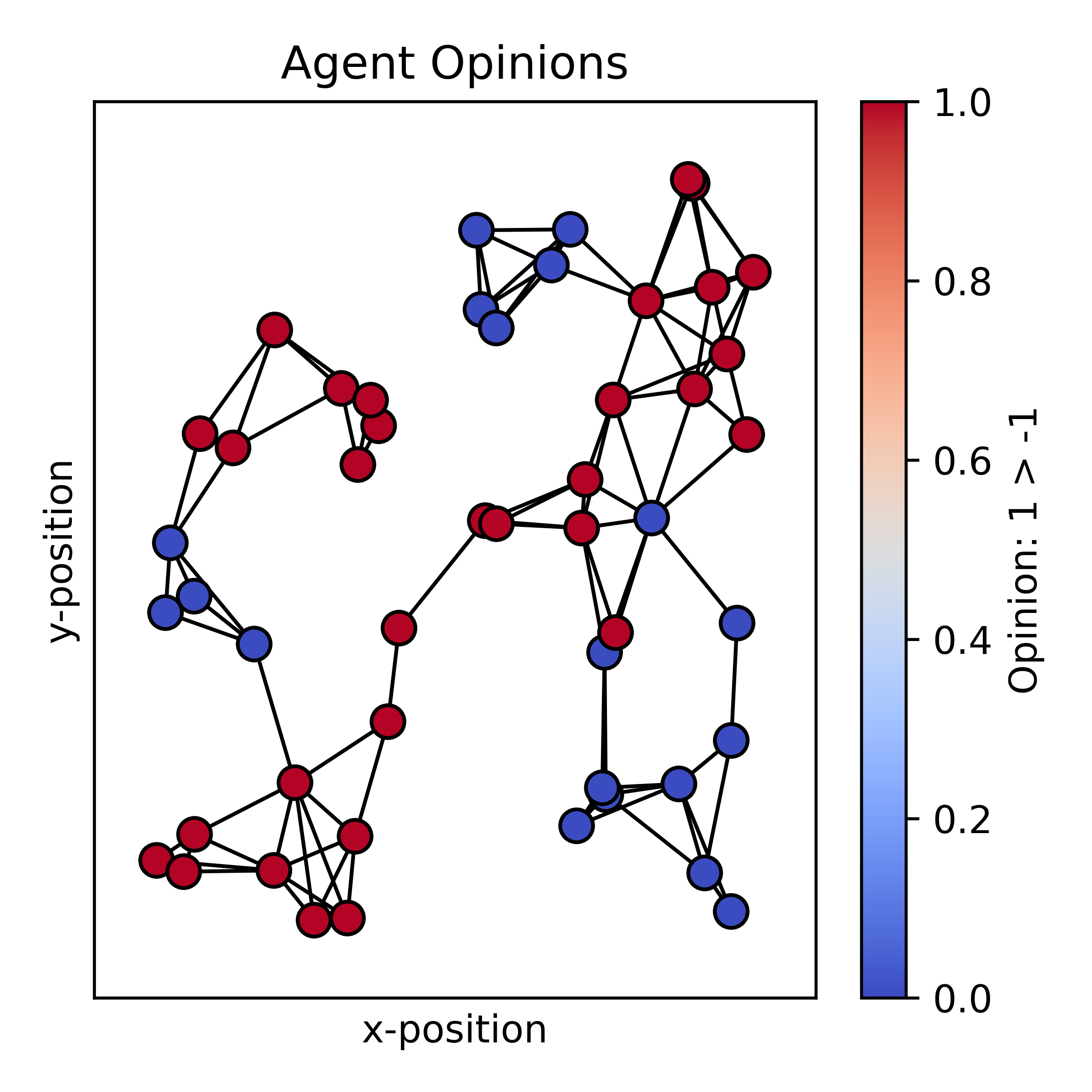}}
\subfloat[]{\includegraphics[trim = {0 0 65 0}, clip = true, width = 0.27\textwidth]{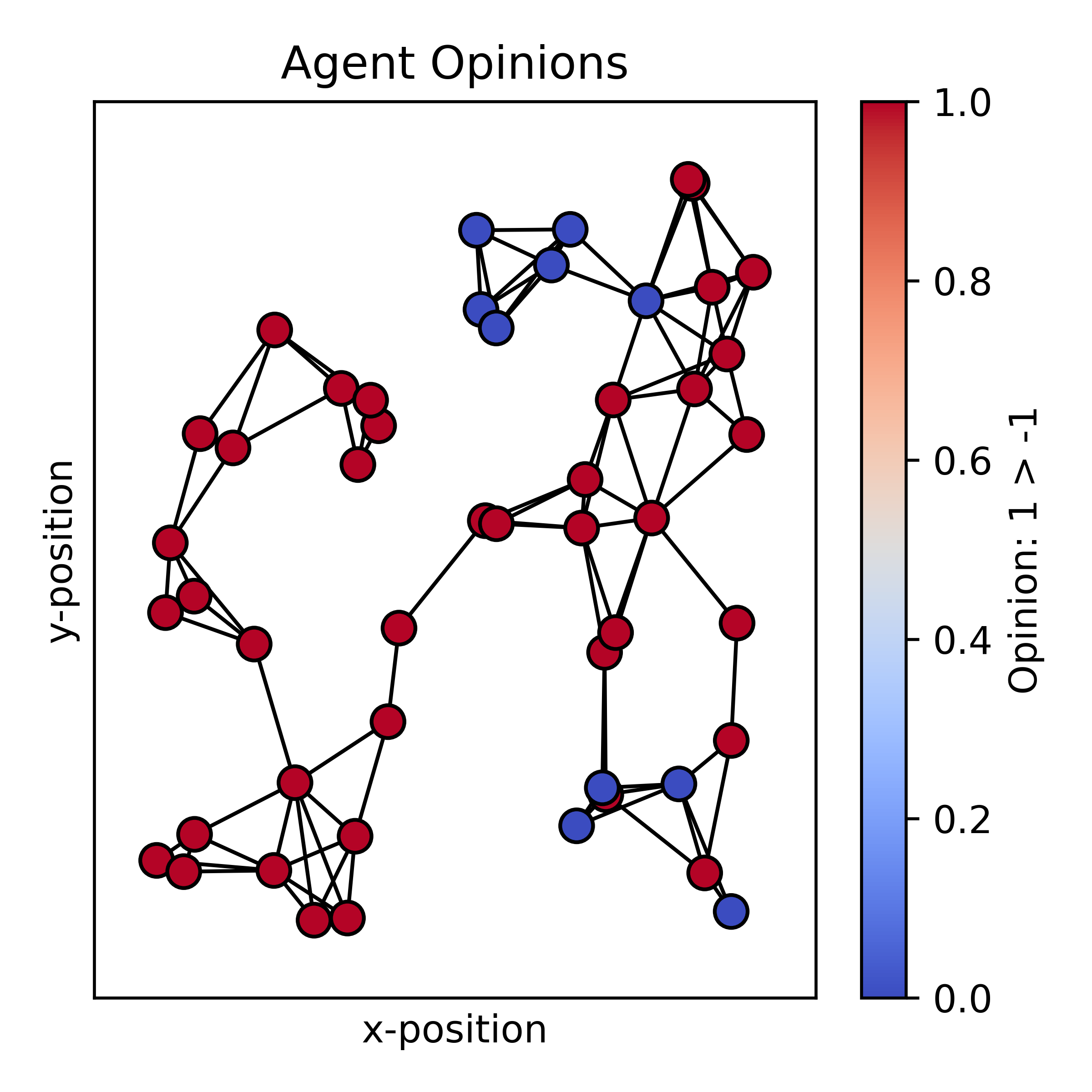}}\\
    \subfloat[]{\includegraphics[trim = {0 0 65 0}, clip = true, width = 0.27\textwidth]{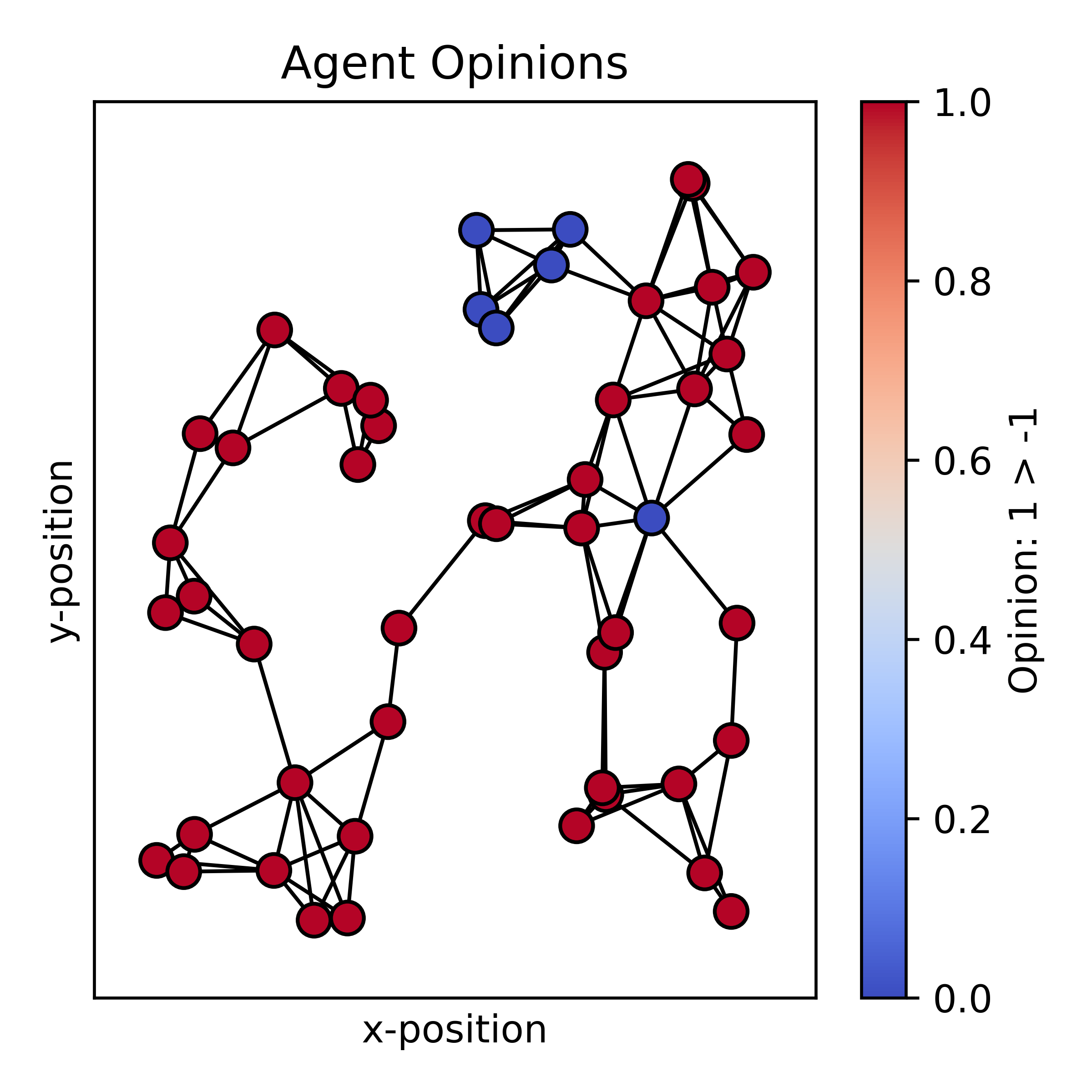}}
\subfloat[]{\includegraphics[trim = {0 0 65 0}, clip = true, width = 0.27\textwidth]{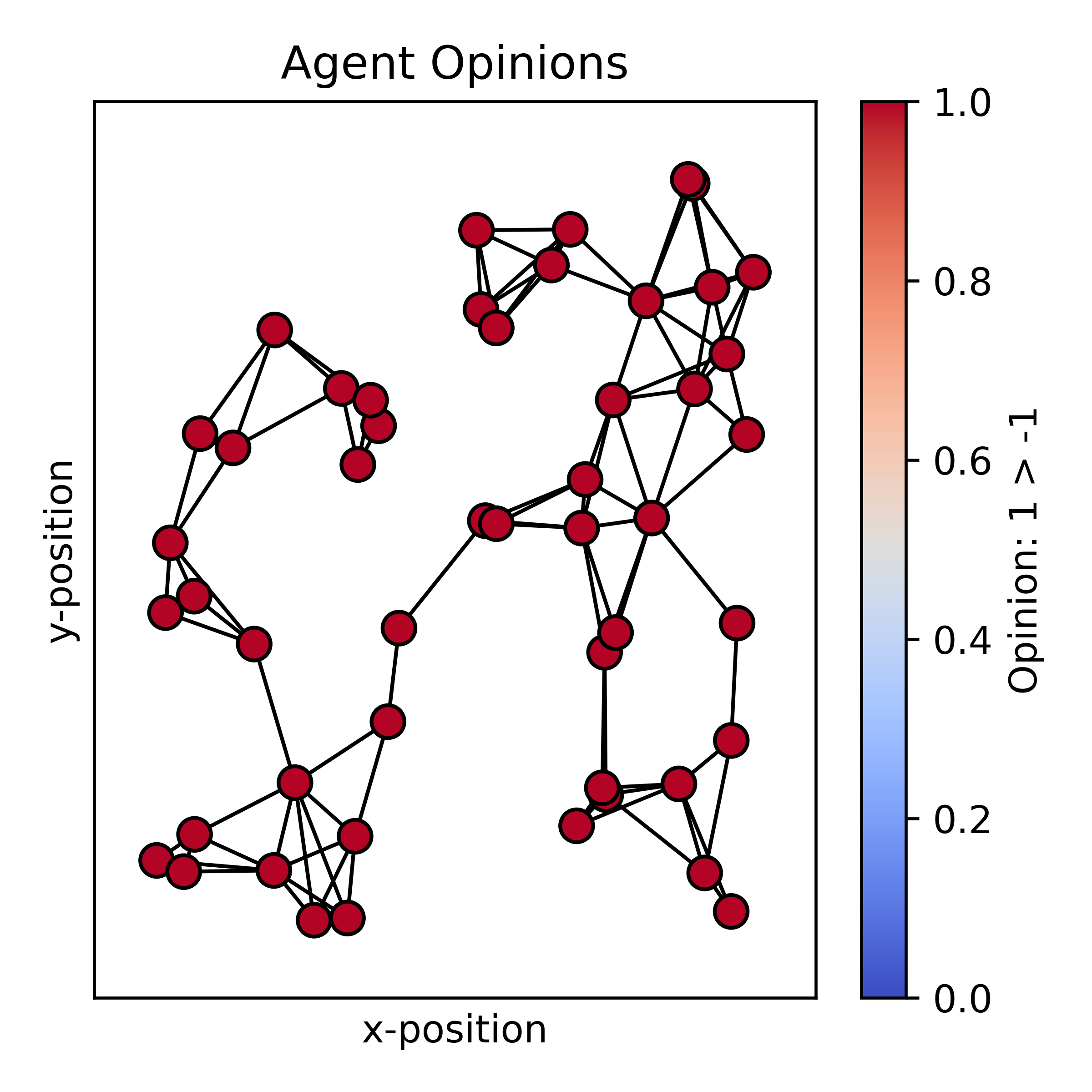}}
\subfloat[]{\includegraphics[trim = {0 0 65 0}, clip = true, width = 0.27\textwidth]{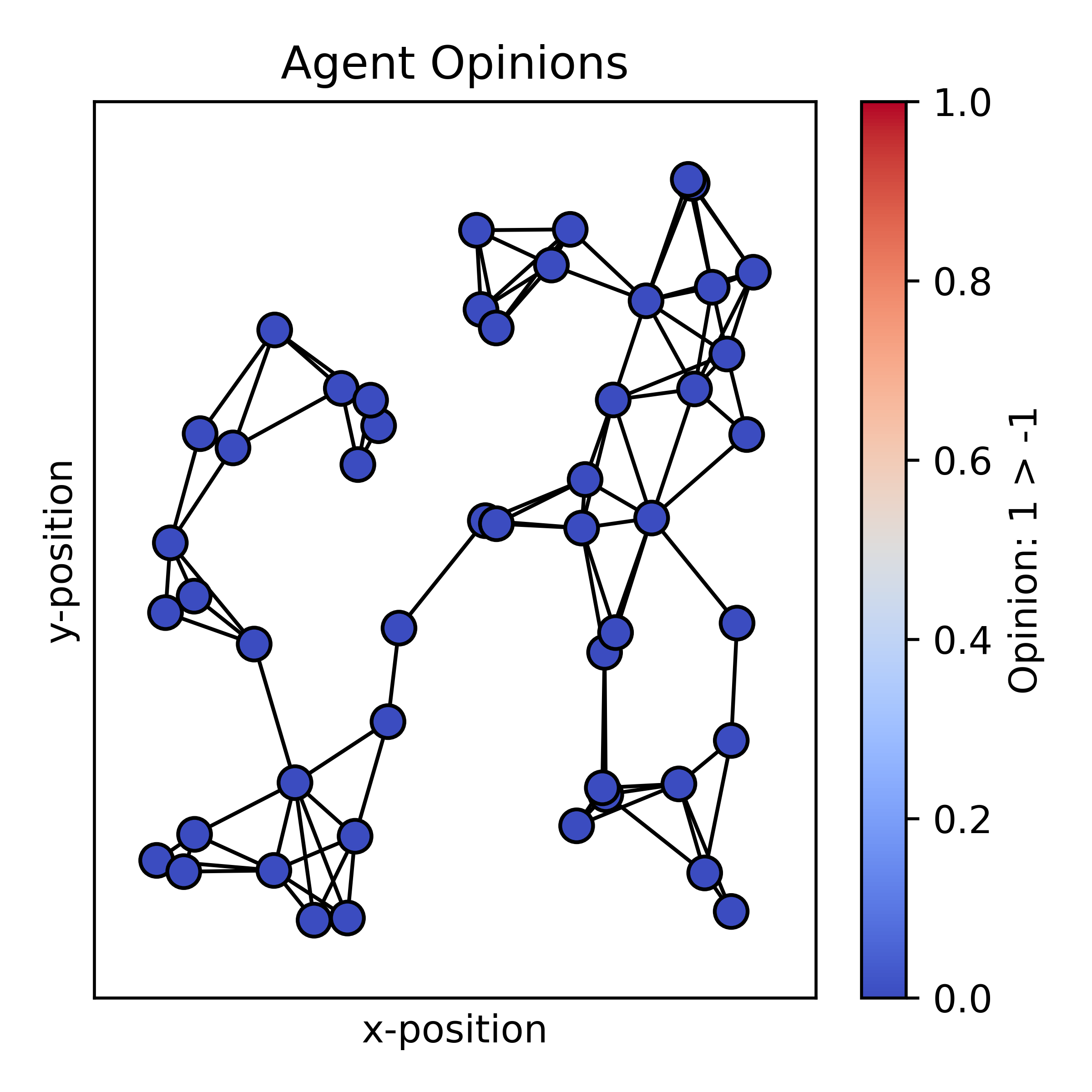}}
    \caption{Opinions in simulation run of the SRLOD model at rounds {(A) $t=1$, (B) $t=10^3$, (C) $t=10^4$, (D) $t=2\times 10^5$, (E) $t=3\times 10^5$, and (F) $t=2\times 10^6$}. Notice the switch from consensus on opinion $o=1$ {(red)} to opinion $o=-1$ {(blue)} between $t=3\times 10^5$ and $t=2\times 10^6$.}
    \label{fig:erg_nets}
\end{figure}

The dynamics of the SRLOD model are ergodic by Proposition~\ref{prop:erg}, meaning that the process can reach all states from all other states. In the ergodic setting on a finite state space, one can look for a stationary distribution. That is a probability distribution over the states reporting the probability of observing each state as $t\to \infty$. Thus, the questions one might ask of the model changes from `What is the probability of consensus?' to `What proportion of time does the system spend in each state?' To illustrate the ergodicity of the SRLOD model, we show a simulation run in which consensus was reached on both opinions in Figures~\ref{fig:erg_total_log} and~\ref{fig:erg_nets}. 

In this simulation we set $\alpha = 0.25$, $\epsilon = 0.2$, $N=50$ and $r_g = 0.2$. Because consensus is no longer stable, we stop the simulation manually at $t=8\times 10^6$. The number of agents holding opinion $o=1$ is plotted on a logarithmic timescale in Figure~\ref{fig:erg_total_log} while the state of the network is shown for telling rounds in Figure~\ref{fig:erg_nets}. {The reason for reducing the number of agents in this simulation is to speed up the dynamics thus making it possible to observe the phenomena in a relatively short simulation.}

\end{document}